\documentclass[reprint,amsmath,amssymb,
nobibnotes,
aps,twocolumn,prb]{revtex4-1}
\usepackage{graphicx}% Include figure files
\usepackage{dcolumn}% Align table columns on decimal point
\usepackage{bm,braket,bbold}% bold math
\usepackage{natbib}
\usepackage{hyperref}
\hypersetup{colorlinks=true,citecolor=blue,linkcolor=red,urlcolor=blue} 
\usepackage{float,color}
\usepackage{epsfig}
\usepackage{mwe}
\usepackage{bbm}
\graphicspath{{images/}}
\begin{document}

\preprint{APS/123-QED}

\title{Theory of non-Markovian dynamics in resonance fluorescence spectrum}% Force line breaks with \\

\author{Abhishek Kumar${}^{1,2,3}$}
\email{ak@csrc.ac.cn}
\affiliation{${}^1$Beijing Computational Science Research Center, Beijing 100193, China\\
${}^2$School of Science and Engineering, Reykjavik University, Menntavegi 1, IS-101 Reykjavik, Iceland\\
${}^3$Department of Physics, McGill University, Montr\'{e}al, Qu\'{e}bec H3A 2T8, Canada
}

\date{\today}% It is always \today, today,

\begin{abstract} 
We present a detailed theoretical study of non-Markovian dynamics in the fluorescence spectrum of a driven semiconductor quantum dot (QD), embedded in a cavity and coupled to a three-dimensional (3D) acoustic phonon reservoir. In particular, we investigate the effect of pure dephasing on one of the side-peaks of the Mollow-triplet spectrum, expressed in terms of the off-diagonal element of the reduced system operator. The QD is modeled as a two-level system with an excited state representing a single exciton, and ground state represents the absence of an exciton. Coupling to the radiative modes of the cavity is treated within usual Born-Markov approximation, whereas dot-phonon coupling is discussed within non-Markovian regime beyond Born approximation. Using an equation-of-motion technique, the dot-phonon coupling is solved exactly and found that the exact result coincides with that of obtained within Born approximation. Furthermore, a Markov approximation is carried out with respect to the phonon interaction and compared with the non-Markovian lineshape for different values of the phonon bath temperature. We have found that coupling to the phonons vanishes for a resonant pump laser. For a non-resonant pump, we have characterzied the effect of dot-laser detuning and temperature of the phonon bath on the lineshape. The sideband undergoes a distinct narrowing and aquires an asymmetric shape with increasing phonon bath temperature. We have explained this behavior using a dressed-state picture of the QD levels.
%\begin{description}
%\end{description}
\end{abstract}

\pacs{Valid PACS appear here}% PACS, the Physics and Astronomy
\maketitle

%\tableofcontents

\section{\label{sec:introduction}Introduction}
The laws of quantum mechanics allow for quantum computers\cite{vonneumann55,benioff80,feynman82}, which are known to be significantly more powerful than classical computers. In a quantum computer, information is stored in quantum bits (qubits), rather than classical bits\cite{loss97,schumacher95}. A single qubit represents a zero or one and is a two-level system with two energy levels which can be used to store and process the information\cite{loss98}. An example of a two-level system, frequently used in quantum optics, is composed of the ground and excited states of an atom. Semiconductor QDs can be modeled as a two-level system with one exciton in the excited state\cite{zrenner02,brunner09,strauss17}. Semiconductor QDs embedded inside a cavity has been a subject of intense research as a promising candidate for quantum computation and information tasks, as well as a source of single photon\cite{warburton15}.

Recently, experiments on the cavity embedded QDs have been reported to show different spectral features of the Mollow-triplet fluorescence spectrum\cite{Ulhaq13,cui06,dpsm10}. In particular, a dot coupled to the acoustic phonon bath in the super-ohmic regime has shown modified spectral features as a function of the phonon bath temperature. More precisely, the triplet sideband is observed to show a systematic spectral sideband broadening for both resonant and off-resonant cases. This problem was studies experimentally\cite{ulrich11prl} and anlyzed theoretically\cite{hughes12prb} in terms of the usual Born-Markov approximation. However, the pure dephasing process a 3D phonon bath is in the form a super-ohmic independent boson model (IBM), which is known to be highly non-Markovian\cite{dpsm13prl,weiler12,krummheuer02,vagov14}, and these results have been studied and discussed in terms of usual Born-Markov approximation\cite{agarwal17,shelykh17,hughes12prb}. 

The system correlation function, for non-Markovian interactions (for e.g. nuclear spins\cite{bill04}, phonons\cite{kaer10}), decays with a typical time scale given by the correlation time $\tau_{c}$ which never dies-off to zero. In other words, the correlation time is non-zero and can be larger than the system decay time $\tau_{S}$, which is the signature of a strongly history-dependent non-Markovian interaction. 

Furthermore, the equation of motion for correlation function has an additional term known as irrelevant part, which is non-zero for non-Markovian interactions. This additional term vanishes for a Markov approximation, when the well-known QRT can be applied to find the system correlation\cite{swan81}. Recent theories [for e.g. Ref.~\onlinecite{hughes12prb}] discussing fluorescence spectrum in solid-state systems rely on a history-independent Markov process and apply the usual QRT, giving rise to an exponential decay of the system correlation. Often, physical processes in solid-state systems\cite{thanopulos17,toyli16,dpsm16} are highly non-Markovian (history-dependent), and so the resulting spectrum using the QRT can no longer be used to describe their spectral properties.

In this paper, we analyze the effect of pure dephasing due to a 3D acoustic phonon bath on the fluorescence spectrum of a cavity coupled to a semiconductor QD. The associated emission spectrum can be directly related to a correlation function for system observables, which we evaluate beyond the Markov approximation using a Nakajima-Zwanzig GME\cite{fick,swan81}. Assuming a large band-width, coupling to the radiation modes of the cavity can be treated within Born-Markov approximation, which is relevant to cavity quantum electrodynamics (cavity-QED) experiments\cite{muller07}. The dot is represented by a two-level system with an exciton in the excited state coupled to a phonon reservoir, and can be modeled with the usual IBM\cite{mahan}. 

The resultant fluorescence spectrum has three components due to dressing of the levels by a pump laser\cite{Mollow69,cohen}, and we project the system into a dressed state basis which allows us to characterize the three components of triplet separately\cite{hopfmann17,laucht17}. We have solved the dot-phonon coupling using an exact approach beyond Born approximation, and the exact result coincides with that of obtained within Born approximation. Phonon coupling gives rise to a frequency dependent frequent-shift and dephasing, which bring non-Lorentzian features in the fluorescence spectrum. Frequency-shift and dephasing due to phonon interaction are strongly temperature-dependent and vanish for the lower temperatures. We found that in the dressed-state basis levels of interest are coupled asymmetrically to the phonons and have vanishing dephasing and frequency-shift for a resonant dot-laser frequency. We have also observed that the sideband undergoes a distinct narrowing and becomes asymmetric with increasing temperature, which is explained using the dressed-states energy levels.

This paper is organized as follows: In Sec.~\ref{sec:fluorescence}, we start with discussing the setup and establish the formula for resonance fluorescence spectrum of a general two-level system. In Sec.~\ref{sec:model}, we introduce the model Hamiltonian for a driven cavity-QED two-level system interacting with a phonon bath. In Sec.~\ref{sec:gme}, we discuss and derive the exact form of Nakajima-Zwanzig GME for the dynamics of the reduced density matrix and correlation function. We obtain the expressions for the lineshape functions in both Markovian and non-Markovian regimes. In Sec.~\ref{sec:results}, we present our results and discuss the plots in different parametric regimes. In Sec.~\ref{sec:conclusions}, we conclude with a discussion and summary of the results. Other technical details are discussed in the Appendixes.

\section{\label{sec:fluorescence}Fluorescence spectrum}
\begin{figure}[ht!]
\includegraphics[width=0.5\textwidth]{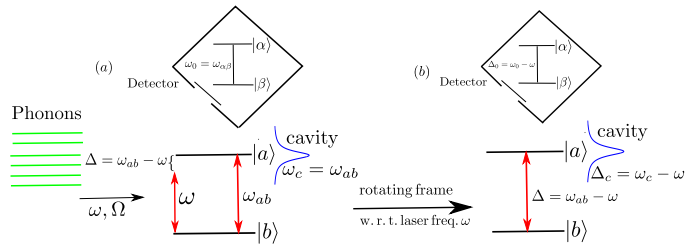}
\caption{(Color online)  (a) A two-level system inside a cavity with natural frequency $\omega_{ab}$ is driven by a laser pump frequency $\omega$. A phonon bath, shown by the green lines, is coupled to the excited state $\ket{a}$ of the QD. (b) In the rotated frame with respect to the pump laser, where $\Delta=\omega_{ab}-\omega\,(\Delta_c=\omega_{c}-\omega, \Delta_0=\omega_{0}-\omega)$ is detuning of the transition (cavity, probe) frequency from the laser. Here, role of the spectrum analyzer is to admit frequencies which are resonant with the transition frequency of the detector. The photonic density of states of the cavity is described by a Lorentzian spectrum, and $\omega_{c}$ is the central frequency of the cavity mode.}
\label{fig:tls:rwa}
\end{figure}

We start with the model Hamiltonian of a general two-level system interacting with radiation modes of the electromagnetic field, which can be written as system ($H_S$), field ($H_R$), and interaction ($H_{SR}$) in terms of the standard Jaynes-Cummings model within a rotating-wave approximation (RWA):
\begin{align}
H_{0}&=H_{S}+H_{R}+H_{SR},\\
H_{R}&=\sum_{k}\omega_{k}a_{k}^{\dagger}a_{k},\\
H_{SR}&=\sum_{k} g_{k}(\sigma_{ab}a_{k}+\sigma_{ba}a_{k}^{\dagger}),
\end{align}
where  $\sigma_{ab}=|a\rangle\langle b|$ and $\sigma_{ba}=|b\rangle\langle a|$ are the raising and lowering operators between excited state $|a\rangle$ and ground state $|b\rangle$ in the Hilbert space of the system and $a_{k}$, $a_{k}^{\dagger}$ are the annihilation and creation operators in the Hilbert space of a set of electromagnetic modes coupled to the system. Coupling to the radiation modes is given by a coupling constant $g_{k}$ to the mode of frequency $\omega_{k}$. For simplicity of the units, we have also used $\hbar=1$. Here, we adopt the well-known theory of gedanken spectrum analyzer given in Ref.~[\onlinecite{scully}], and assume that the radiation field emitted by the system is detected by a two-level atom (detector) with a transition frequency $\omega_{\alpha}-\omega_{\beta}=\omega_{0}$, which is prepared in its ground state $\ket{\beta}$ initially, see Fig.~\ref{fig:tls:rwa}.

The Hamiltonian for the detector is given by
\begin{equation}
H_{D}=\frac{\omega_{0}}{2}\left(|\alpha\rangle\langle\alpha|-|\beta\rangle\langle\beta|\right),
\end{equation}
and coupling between detector atom $H_{D}$ and the radiation field, included in the Hamiltonian $H$, is given by the Hamiltonian 
\begin{equation}
H_{DR}=\sum_{k} g_{k}^{D}\left(|\alpha\rangle\langle\beta|a_{k}+|\beta\rangle\langle\alpha|a_{k}^{\dagger}\right),
\end{equation}
where $g_{k}^{D}$ is the coupling of detector to a field mode $k$. Therefore, the Hamiltonian for entire system including system and detector, as well as coupling to the radiation modes can be written as:
\begin{equation}
H=H_{0}+H_{D}+H_{DR}.
\end{equation}
According to Wiener-Khintchine theorem, fluorescence spectrum, in the stationary regime and in the interaction picture with respect to the detector, is given by Fourier transform of the correlation function\cite{scully}
\begin{equation}
S(\omega_{0})=|\mathcal{\wp_{\alpha\beta}}|^{2}\,\mathrm{Re}\int_{0}^{\infty}d\tau\,\langle E^{(-)}(0)E^{(+)}(\tau)\rangle\,e^{i\omega_{0}\tau},
\label{eqn:flou:spec:1}
\end{equation}
where $\wp_{\alpha\beta}$ is the detector dipole matrix element with positive-frequency part of the electric field is defined by 
\begin{equation}
E^{(+)}(t) =\sum_{k}\varepsilon_{k}a_{k}(t),
\end{equation}
and the negative-frequency part of the electric field is $E^{(-)}(t)=[E^{(+)}(t)]^{\dag}$. The quantity $\varepsilon_{k}=\sqrt{\hbar\omega_{k}/(2\epsilon_{0}V)}$ is the electric field per photon and $V$ is the effective volume of a cubic cavity resonator. We rewrite the correlation function, in Eq.~(\ref{eqn:flou:spec:1}), in terms of the system operators using the methods described in Refs.~\onlinecite{scully,cohen}:
\begin{equation}
S(\omega_{0})=\bar{I}^{2}\,\mathrm{Re}\int_{0}^{\infty}d\tau\,\langle \sigma_{ab}(0)\sigma_{ba}(\tau)\rangle\,e^{i\omega_{0}\tau},
\label{eqn:flou:spec:11}
\end{equation} 
here $\bar{I}$ is the detector response function [discussed in Appendix \ref{sec:detector response}]. The average $\langle\dots\rangle=Tr\lbrace\dots\bar{\rho}\rbrace$ , in Eq.~(\ref{eqn:flou:spec:11}), is given with respect to the stationary density matrix $\bar{\rho}$, where $\bar{\rho}=\lim\limits_{T\to\infty}\frac{1}{T}\int_{0}^{T}dt\,\rho(t)$. Using the cyclic property of trace, fluorescence spectrum can be written as
\begin{equation}
S(\omega_{0})=\bar{I}^{2}\,\mathrm{Re}\int_{0}^{\infty}d\tau\,Tr\lbrace\sigma_{ba}\Omega(t)\rbrace\,e^{i\omega_{0}t},
\label{eqn:flou:spec:12}
\end{equation}
here operator $\Omega(t)$ is defined as 
\begin{equation}
\Omega(t)=e^{-iH_{0}t}\bar{\rho}\sigma_{ab}e^{iH_{0}t},
\label{eqn:full operator}
\end{equation} 
and $H_{0}=H_{S}+H_{R}+H_{SR}$. Since $\sigma_{ab}$ and $\sigma_{ba}$ are operators in the system Hilbert space and $[H_{D},H_{0}]=0$, the evolution of $\Omega(t)$ is determined by the Hamiltonian of emitting system and radiation field, $H_{0}$, in the absence of detector and can be computed without using the well-known QRT\cite{swan81}.

\section{\label{sec:model}Model}
\subsection{\label{sec:hamiltonian}Hamiltonian}
We consider a driven two-level cavity-QED system with an excited state $\ket{a}$ representing a single exciton, and a ground state $\ket{b}$ with no exciton. The QD interacts with the cavity photons and a phonon reservoir which is coupled to the excited state $\ket{a}$, as shown in Fig.~\ref{fig:tls:rwa}. The Hamiltonian for the total system reads,
\begin{align}
H(t) =&\frac{\omega_{ab}}{2}\sigma_{z}+\frac{\Omega}{2}(\sigma_{ab}+\sigma_{ba})(e^{i\omega t}+e^{-i\omega t})\nonumber\\
&+\sum_{k}\omega_{k}a_{k}^{\dagger}a_{k}
+\sum_{k}g_{k}(\sigma_{ab}+\sigma_{ba})(a_{k}+a_{k}^{\dagger})\nonumber\\
&+\sum_{k}\omega_{q}b_{q}^{\dagger}b_{q}+\sum_{q}\lambda_{q}\,\sigma_{aa}(b_{q}+b_{q}^{\dagger}),
\label{eqn:full pTLS H}
\end{align}
where $\omega_{ab}$ is the transition frequency of the two-level system and $\omega$ is the frequency of laser field. The photon (phonon) modes are represented by bosonic fields with frequencies $\omega_{k}$ ($\omega_{q}$) with creation and annihilation operators $a_{k}^{\dagger}$ ($b_{q}^{\dagger}$) and $a_{k}$ ($b_{q}$), respectively. The system-photon (system-phonon) coupling strength is given by $g_{k}$ ($\lambda_{q}$), and $\Omega$ (Rabi frequency) is the coupling between the two-level system and laser field. The system operators are denoted by $\sigma_{ij}=|i\rangle\langle j|$ where $i,j\in \lbrace a,b\rbrace$ and $\sigma_{z}=\sigma_{aa}-\sigma_{bb}$.

The explicit time dependence in $H(t)$ [Eq.\ (\ref{eqn:full pTLS H})] can be removed by going to a rotating frame and applying a RWA. We perform the RWA on both the driving term and the system-photon coupling term, in which we neglect the rapidly-oscillating terms and keeping only the time-independent part. The resulting RWA Hamiltonian in the rotated frame is then
\begin{align}
\tilde{H}=&\frac{\Delta}{2}\sigma_{z}+\frac{\Omega}{2}(\sigma_{ab}+\sigma_{ba})+\sum_{k}\Delta_{k}a_{k}^{\dagger}a_{k}\nonumber\\
&+\sum_{k}g_{k}(\sigma_{ab}a_{k}+\sigma_{ba}a_{k}^{\dagger})\nonumber\\
&+\sum_{k}\omega_{q}b_{q}^{\dagger}b_{q}+\sum_{k}\lambda_{q}\,\sigma_{aa}(b_{q}+b_{q}^{\dagger}),
\end{align}
where $\Delta=\omega_{ab}-\omega$ and $\Delta_{k}=\omega_{k}-\omega$ are the detunings of atomic and cavity frequencies from the laser pump frequency $\omega$. In general, operators in the rest frame are transformed to the rotating frame according to the following relations:
\begin{align}
\sigma_{z}(t)&=\tilde{\sigma}_{z}(t),\\
\sigma_{ab}(t)&=e^{-i\omega t}\tilde{\sigma}_{ab}(t).
\label{eqn:operator in rotated frame}
\end{align}
Due to presence of the intense laser field the two bare states are strongly coupled to each other and give rise to two dressed states. The dressed states can be written, in terms of the bare states, as:
\begin{align}
|+\rangle&={\bf c}|a\rangle+{\bf s}|b\rangle, \label{eqn:dressed plus states}\\
|-\rangle&=-{\bf s}|a\rangle+{\bf c}|b\rangle, \label{eqn:dressed minus states}
\end{align}
where ${\bf c}=\cos\theta$ and ${\bf s}=\sin\theta$ with the mixing angle $\theta$ is given by $\tan\theta=\sqrt{(\Omega_{R}-\Delta)/(\Omega_{R}+\Delta)}$, where $\Omega_{R}=\sqrt{\Omega^2+\Delta^2}$ is the dressed Rabi frequency.
\begin{figure}[htb]
\centering
\includegraphics[width=0.45\textwidth]{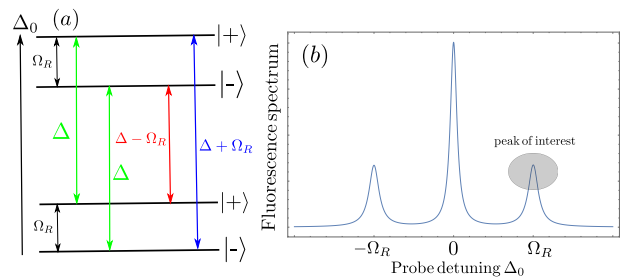}
\caption{(Color online) (a) Energy-level diagram showing the allowed transitions among the dressed state energy-levels, which give rise to the Stokes (red line) and anti-Stokes (blue line) sidebands along with the central peak (green line) of the Mollow triplet spectrum. (b) Schematic plot of the fluorescence spectrum. The satellite peak of interest is centered at $\Delta_{0}=\Omega_{R}$.}
\label{fig:dressed-states}
\end{figure}
Furthermore, following closely the discussing on IBM in Ref.~[\onlinecite{mahan}], we apply another transformation by using a canonical transformation $H'=e^{B}\tilde{H}e^{-B}$,
where $B=\frac{1}{2}\mathbb{1}\otimes\sum_{q}\frac{\lambda_{q}}{\omega_{q}}(b_{q}^{\dagger}-b_{q})$ is an anti-hermitian operator, to write the total Hamiltonian in terms of the free and perturbed parts as, after transforming it to the dressed-state basis
\begin{equation}
\bar{H}\simeq H_{0}+H_{V},\label{eqn:polaron:hamiltonian}
\end{equation} 
where the free part is
\begin{align}
H_{0}&=H_{S}+H_{R}+H_{P},\\
H_{S}&=\frac{\Omega'_{R}}{2}\sigma_{3},\\
H_{R}&=\sum_{k}\Delta_{k}a_{k}^{\dag}a_{k},\\
H_{P}&=\sum_{q}\omega_{q}b_{q}^{\dag}b_{q}
\end{align}
and the perturbed part is given by
\begin{align}
H_{V}&=H_{dR}+H_{SR}+H_{dP},\\
H_{dR}&={\bf c.s}\,\sigma_{3}\,\sum_{k}g_{k}(a_{k}+a_{k}^{\dag}),\\
H_{dP}&=\frac{{\bf c}^{2}-{\bf s}^{2}}{2}\,\sigma_{3}\,\sum_{q}\lambda_{q}(b_{q}+b_{q}^{\dag}),\\
H_{SR}&=\sum_{k}g_{k}\left[({\bf c}^{2} a_{k}-{\bf s}^{2} a_{k}^{\dagger})\sigma_{+-}+h.c.\right]\label{eqn:trans:rad},
\end{align}
where $\sigma_{ij}=|i\rangle\langle j|$, $i,j\in \lbrace +,-\rbrace$ and $\sigma_{3}=|+\rangle\langle +|-|-\rangle\langle -|$. The polaron transformation introduces a frequency shift as: $\Omega_{R}'=\Omega_{R}-\Delta_{P}$, where $\Delta_{P}=({\bf c}^{2}-{\bf s}^{2})\sum_{q}\lambda_{q}^{2}/\omega_{q}$ is the polaron shift. We have also performed a Schrieffer-Wolff transformation on the above Hamiltonian to get rid of the energy exchange term ($T_{1}$ lifetime process) due to phonons, see Appendix~\ref{sec:schrieffer-wolff}. The separation into pure dephasing and transition terms is determined by the form of system coupling, i.e.\ dephasing terms contain the diagonal coupling and transition terms contain the off-diagonal couplings. Accordingly, fluorescence spectrum, in Eq.~(\ref{eqn:flou:spec:12}), can be written in Laplace domain using the dressed-state representation as a three-peak spectrum [Ref.~\onlinecite{cohen}]:
\begin{align}
S(\Delta_{0})=&\bar{I}^{2}\,\mathrm{Re}\bigg[{\bf c.s}\,\Omega_{z}(s)+{\bf c}^{2}\Omega_{+-}(s)\nonumber\\
&-{\bf s}^{2}\Omega_{-+}(s)\bigg]_{s=-i\Delta_{0}},
\label{eqn:three-peak}
\end{align}
where $\Delta_{0}=\omega_{0}-\omega$ is the probe detuning, we have used the relations $\Omega_{ij}=\langle j|\Omega|i\rangle$, $i,j\in \lbrace +,-\rbrace$ and $\Omega_{z}=\Omega_{++}-\Omega_{--}$. The polaron transformation does not affect the fluorescence spectrum in above expression as $B$ acts on the phonon mode Hilbert space and commutes with the system operators. It can be seen that the first term in Eq.~(\ref{eqn:three-peak}) gives the central peak whereas last two terms give rise to satellite peaks of the Mollow-triplet, shown in Fig.~\ref{fig:dressed-states}. We are interested in the influence of pure dephasing due to phonon interaction, which only affects the off-diagonal elements of the system operator. From here and what follows, we will study the effect of pure dephasing on the fluorescence spectrum, with a particular focus on one of the Mollow-triplet sidebands (Stokes line). Alternatively, when the width of each side peak $\sim\Gamma$, [see Eq.~(\ref{eqn:mar:decay:rate}), below] is small compared to the peak separation $\sim\Omega_R$, we approximate the spectrum near the side peak centered at $\Delta_0\simeq\Omega_R$ by, see Fig.~\ref{fig:dressed-states}
\begin{equation}
S(\Delta_0)\simeq S_+(\Delta_0)=\bar{I}^{2}{\bf c}^{2}\,\mathrm{Re}[\Omega_{+-}(s=-i\Delta_{0})].
\label{eqn:one-peak}
\end{equation}
In order to compute the spectrum, we will evaluate the dynamics of matrix element $\Omega_{+-}(t)={\rm Tr}[\sigma_{-+}\Omega(t)]$ in the dressed-state representation with Hamiltonian in the polaron frame, given by Eq.~(\ref{eqn:polaron:hamiltonian}).

\subsection{Initial conditions}
The radiation and phonon modes are decoupled from the system for times $t_{0}<0$ (where $t_0$ is a time in the distant past), and prepared independently in the states described by density matrices
$\rho_{R}(t_{0})$, $\rho_{P}(t_{0})$ and $\rho_{S}(t_{0})$, respectively. The interactions (photons and phonons) are
switched on at this time $t=t_{0}$, and the state of entire system is described by the full density matrix $\rho(t_{0})$:
\begin{equation}
\rho(t_0)=\rho_{R}(t_0)\otimes\rho_{P}(t_0)\otimes\rho_{S}(t_0),
\label{eqn:factorize initial condition}
\end{equation}
where initial density matrix for photon is described by vacuum of the cavity modes 
\begin{equation}
\rho_{R}(t_{0})=\prod_{k}\ket{0_{k}}\bra{0_{k}},
\label{eqn:photon density matrix}
\end{equation}
and phonon modes are described by a canonical ensemble at temperature $T$:
\begin{equation}
\rho_{P}(t_{0})=\frac{\mathrm{exp}(-H_{P}/k_{B}T)}{{\rm Tr}[\mathrm{exp}(-H_{P}/k_{B}T)]}.
\label{eqn:phonon density matrix}
\end{equation}
We recall the operator $\Omega(t)$, where
\begin{equation}
\Omega(t)=e^{-i\bar{H}t}\bar{\rho}\sigma_{ab}e^{i\bar{H}t}
\end{equation}
is analogous to the density matrix operator with a modified initial condition, given by
\begin{equation}
\Omega(0)=\bar{\rho}\sigma_{ab},
\label{eqn:corr:initial:condition}
\end{equation}
where stationary density matrix $\bar{\rho}$ and hence $\Omega(0)$ account for conditions that accumulate between the system and interactions in the time interval $t\in[t_{0} ,0]$. We choose an initial condition when exciton is in excited state $\ket{a}$ given by $\rho_{S}(t_{0})=\ket{a}\bra{a}$, and evolves in the presence of pump laser. We switch on the detector at $t=0$ and subsequently calculate the dynamics of $\Omega_{+-}(t)$ for $t>0$, with initial condition given by the steady state density matrix accumulated between the time interval $[t_{0}, 0]$.

\section{\label{sec:gme}Generalized Master Equation} 
We are interested in the dynamics of reduced system operator, after tracing over variables of photon and phonon modes; $\Omega_{S}(t)={\rm Tr_{R}Tr_{P}}\Omega(t)$. To study the dynamics of reduced system operator $\Omega_{S}(t)$, we introduce a projection superoperator $P$, defined by its action on an operator: $P\mathcal{O}(t)=\rho_{R}(t_0)\rho_{P}(t_0){\rm Tr_{R}Tr_{P}}\mathcal{O}(t)$. 

Both operators, $\rho(t)$ and $\Omega(t)$, follow the same von-Neumann equation, and can be written in form of exact Nakajima-Zwanzig GME\cite{swan81}, using $Q\rho(t_{0})=0$
\begin{equation}
P\dot{\rho}(t)=-iPLP\rho(t)-i\int_{t_{0}}^{t}dt'\ \Sigma(t-t')P\rho(t'),
\label{eqn:GME for density matrix}
\end{equation}
where $\Sigma(t)$ is the self-energy superoperator 
\begin{equation}
\Sigma(t)=-iPLQ\,e^{-iQLt}QLP,
\label{eqn:self-energy superoperator}
\end{equation}
and $L$ is the full Liouvillian superoperator, defined as $L_{\alpha}\mathcal{O}=[H_{\alpha},\mathcal{O}]$ and $\alpha=0(S,R,P),V(dR,dP,SR)$. We have used the properties of projection operator: $P^{2}=P$ and
\begin{equation}
\langle\mathcal{O}_{S}\rangle(t)=\mathrm{Tr}\lbrace\mathcal{O}_{S}\rho(t)\rbrace=\mathrm{Tr}\lbrace \mathcal{O}_{S}P\rho(t)\rbrace,
\end{equation}
also introducing its complement $Q=\mathbb{1}-P$. We can derive an equation of motion for $\Omega(t)$ analogous to the equation for $\rho(t)$ [Eq. (\ref{eqn:GME for density matrix})]. However, an additional term appears because $Q\Omega(0)=Q\bar{\rho}\sigma_{ab}\neq 0$, and we have assumed that the full density matrix operator is not separable for all times i.e. $\rho(t)\neq\rho_{R}(t_0)\otimes\rho_{P}(t_0)\otimes\rho_{S}(t)$. The resulting motion equation for $P \Omega (t)$ is then,
\begin{align}
P\dot{\Omega}(t)=&-iPLP\Omega(t)-i\int_{0}^{t}dt'\ \Sigma(t-t')P\Omega(t')\nonumber\\
&-iPLQe^{-iQLt}Q\Omega(0),
\label{eqn:GME for pseudo density matrix}
\end{align}
where $\Sigma(t)$ is defined in Eq.\ (\ref{eqn:self-energy superoperator}) and the last term in the above equation contains $Q\Omega(0)$, i.e.\ the irrelevant part of $\Omega(0)$ is non-zero. When the radiation and phonon modes are described by Eqs.~(\ref{eqn:photon density matrix}) and (\ref{eqn:phonon density matrix}), the projection operator $P$ follows some useful identities
\begin{align}
PLP&=L_{S}P=PL_{S},\label{eqn:iden1}\\
PL_{V}P&=0,\label{eqn:iden2}\\
PLQ&=PL_{V},\label{eqn:iden3}\\
QLP&=L_{V}P.\label{eqn:iden4}
\end{align}
We apply above identities (\ref{eqn:iden1})-(\ref{eqn:iden4}), and perform the partial traces on Eqs.~(\ref{eqn:GME for density matrix}) and (\ref{eqn:GME for pseudo density matrix}) over variables of radiation and phonon modes to obtain an equation of reduced system operators,
\begin{widetext}
\begin{align}
\dot{\rho}_{S}(t)&=-iL_{S}\rho_{S}(t)-i\int_{t_0}^{t}dt'\ \Sigma_{S}(t-t')\rho_{S}(t'),\label{eqn:eom:rhos}\\
\dot{\Omega}_{S}(t)&=-iL_{S}\Omega_{S}(t)-i\int_{0}^{t}dt'\ \Sigma_{S}(t-t')\Omega_{S}(t')+\Phi_{S}(t),\label{eqn:eom:omegas}\\
\Sigma_{S}(t)&=-i\mathrm{Tr}_{R}\mathrm{Tr}_{P}\bigg[L_{V}e^{-iQLt}L_{V}\rho_{R}(t_{0})\rho_{P}(t_{0})\bigg],\label{eqn:reduced:self:energy}\\
\Phi_{S}(t)&=-i\mathrm{Tr}_{R}\mathrm{Tr}_{P}\bigg[L_{V}e^{-iQLt}Q\Omega(0)\bigg],\label{eqn:irr:part}
\end{align}
where $\Sigma_{S}(t)$ and $\Phi_{S}(t)$ are reduced self-energy and irrelevant part superoperators, respectively, and $\mathcal{O}_{S}(t)=\mathrm{Tr}_{R}\mathrm{Tr}_{P}\mathcal{O}(t)=\sum_{\alpha\beta\in\lbrace +,-\rbrace}\mathcal{O}_{\alpha\beta}(t)\ket{\alpha}\bra{\beta}$ is the reduced system operator. 
\end{widetext}
Comparing Eqs.~(\ref{eqn:eom:rhos}) and (\ref{eqn:eom:omegas}), we found that the first two terms are identical but there is an additional irrelevant part, $\Phi_{S}(t)$, is present in the equation for $\Omega_{S}(t)$. This term vanishes in a Markov approximation and usual QRT can be applied to find the system correlation\cite{swan81}. In addition to this, we have also assumed that the full density matrix is \emph{not} separable for all times which is another reason that QRT is no longer valid in the present case. Furthermore, for a non-Markovian equation the irrelevant term is non-zero and usual QRT can not be used to compute the system correlation in this case\cite{vega08,lax00,jin16}. The equation for off-diagonal matrix element, $\Omega_{+-}$, is coupled to both diagonal $\Omega_{++},\,\Omega_{--}$ and off-diagonal $\Omega_{-+}$ elements of the reduced system operator, and can be written in the form
\begin{widetext}
\begin{align}
\dot{\Omega}_{+-}(t)&=-i\Omega_{R}'\Omega_{+-}(t)-i\int_{0}^{t}dt'\,\Sigma_{+-,+-}(t-t')\Omega_{+-}(t')-i\int_{0}^{t}dt'\,\Sigma_{+-,-+}^{SR}(t-t')\Omega_{-+}(t')\nonumber\\
&-i\int_{0}^{t}dt'\,\Sigma_{+-,++}^{SR}(t-t')\Omega_{++}(t')-i\int_{0}^{t}dt'\,\Sigma_{+-,--}^{SR}(t-t')\Omega_{--}(t')+G_{+-,+-}(t)\Omega_{+-}(0),
\label{eqn:full:eom}
\end{align}
with a non-zero irrelevant part matrix element expressed as (see Appendix \ref{sec:irrelevant-part})
\begin{equation}
[G_{S}(t)]_{+-,+-}=\bigg[-Tr_{R}Tr_{P}L_{V}\,e^{-iQLt}\bigg(\frac{1}{0^{+}+iQL}L_{V}\bigg)\rho_{R}(t_0)\rho_{P}(t_0)\bigg]_{+-,+-}.
\end{equation}
\end{widetext}
The self-energy superoperator can be decomposed into three terms as
\begin{equation}
\Sigma_{S}(t)=\Sigma_{S}^{dR}(t)+\Sigma_{S}^{dP}(t)+\Sigma_{S}^{SR}(t),
\label{eqn:three sigmas}
\end{equation}
where the first and second terms in above expression are pure dephasing ($T_{2}^{*}$ pure dephasing time) processes due to photon and phonon couplings, respectively, and the third term gives rise to transition ($T_{1}$ lifetime) due to radiative modes coupling. We treat the dot-cavity coupling self-energy within Born-Markov approximation to second-order in perturbation Liouvillain $L_{V}$, see Appendix \ref{sec:photon-self-energy}. Applying the continuum of modes for cavity given by Lorentzian density of state [Eq.~(\ref{eqn:dos:cavity})], we find the self-energy matrix elements in their Laplace domian, defined as $f(s)=\int_{0}^{\infty}e^{-st}f(t)dt$,
\begin{align}
\Sigma_{+-,+-}^{dR}(s)\simeq&\frac{-ig^{2}(\Omega_{R}^{2}-\Delta^{2})}{2\Omega_{R}^{2}} \bigg(\frac{1}{s+i(\Omega_{R}'-\Delta_{c})+\Gamma_{c}}\nonumber\\
&+\frac{1}{s+i(\Omega'_{R}+\Delta_{c})+\Gamma_{c}}\bigg),
\label{eqn:dp:pm}
\end{align}
\begin{equation}
\Sigma_{+-,+-}^{SR}(s)\simeq\frac{-ig^{2}}{4\Omega_{R}^{2}}\bigg(\frac{(\Omega_{R}+\Delta)^{2}}{s+i\Delta_{c}+\Gamma_{c}}+\frac{(\Omega_{R}-\Delta)^{2}}{s-i\Delta_{c}+\Gamma_{c}}\bigg),
\label{eqn:trans:pm}
\end{equation}
where $\Delta_{c}=\omega_{c}-\omega$ is the detuning of cavity from laser pump frequency and $\Gamma_{c}$ is cavity bandwidth. Here, Born approximation is justified by finding that the higher order terms in reduced self-energy are suppressed by a small parameter $\sim g^{2}/(\Omega'_{R}\Gamma_{c})$. Similarly for phonon modes, applying a continuum of modes [see Appendix~\ref{sec:phonon-self-energy}] for the deformation potential coupling mechanism\cite{krummheuer02} $N(\epsilon)|\lambda(\epsilon)|^{2}=\alpha_{P}|\epsilon|^{3}e^{-|\epsilon|/\epsilon_{c}}$, where $\alpha_{P}$ is the phonon coupling parameter in the units of ${\emph\rm frequency}^{-2}$ and $\epsilon_{c}$ is the phonon cut-off frequency, we obtain
\begin{align}
\Sigma_{+-,+-}^{dP}(s)=&\frac{-i\alpha_{P}\Delta^{2}}{2\Omega_{R}^{2}}\int_{0}^{\infty}d\epsilon|\epsilon|^{3}e^{\frac{-|\epsilon|}{\epsilon_{c}}}(2n_{B}(\epsilon)+1)\nonumber\\
&\left(\frac{1}{s+i(\Omega'_{R}-\epsilon)}+\frac{1}{s+i(\Omega'_{R}+\epsilon)}\right),
\label{eqn:phonon:self:energy}
\end{align}
here $n_{B}(\epsilon)$ is Bose function. In above expression, the dot-phonon interaction self-energy is evaluated using exact approach within non-Markovian limit and it is found that Born approximation is exact in this case, see Appendix~\ref{sec:exact:self:energy}. Furthermore, the reduced self-energy matrix elements couple to the populations can as well be written within usual Born approximation 
\begin{align}
\Sigma_{++,++}^{SR}(s)\simeq&\frac{-ig^{2}(\Omega_{R}+\Delta)^{2}}{4\Omega_{R}^{2}}\bigg(\frac{1}{s+i(\Omega_{R}'-\Delta_{c})+\Gamma_{c}}\nonumber\\
&+\frac{1}{s-i(\Omega_{R}'-\Delta_{c})+\Gamma_{c}}\bigg)\label{eqn:trans:pp},
\end{align}
\begin{align}
\Sigma_{++,--}^{SR}(s)\simeq&\frac{ig^{2}(\Omega_{R}-\Delta)^{2}}{4\Omega_{R}^{2}}\bigg(\frac{1}{s+i(\Omega_{R}'+\Delta_{c})+\Gamma_{c}}\nonumber\\
&+\frac{1}{s-i(\Omega_{R}'+\Delta_{c})+\Gamma_{c}}\bigg)\label{eqn:trans:mm},
\end{align}
and similarly for the coherence $\Omega_{-+}$,
\begin{align}
\Sigma_{+-,-+}^{SR}(s)\simeq&\frac{-ig^{2}(\Omega_{R}^{2}-\Delta^{2})}{4\Omega_{R}^{2}}\bigg(\frac{1}{s+i\Delta_{c}+\Gamma_{c}}\nonumber\\
&+\frac{1}{s-i\Delta_{c}+\Gamma_{c}}\bigg).
\label{eqn:trans:mp}
\end{align}
Equation for the coherence in Eq.~(\ref{eqn:full:eom}) contains both diagonal and off-diagonal elements of the self-energy superoperator, some of them are fast moving compare to others. In next section, we will perform a secular approximation\cite{cohen} to get rid of the fast oscillating terms.

\subsection{\label{sec:secular:approximation}Secular approximation}
The secular approximation consists in neglecting the fast oscillating terms in Markov equation-of-motion, and the equation for $\Omega_{+-}$ [Eq.~(\ref{eqn:full:eom})] is decoupled from populations and coherence within a secular approximation\cite{cohen}. Here, we consider a general equation of motion for the operator $\Omega(t)$, without irrelevant part matrix elements:
\begin{align}
\dot{\Omega}_{+-}(t)=&-i\Omega_{R}'\Omega_{+-}(t)-i\int_{0}^{t}dt'\,\Sigma_{+-,+-}(t-t')\Omega_{+-}(t')\nonumber\\
&-i\int_{0}^{t}dt'\Sigma_{+-,-+}^{SR}(t-t')\Omega_{-+}(t'),
\end{align}
and would like to perform the secular approximation in order to get rid of the fast oscillating terms. To this end, introducing a rotating frame
\begin{equation}
\Omega'_{+-}(t)=e^{i(\Omega_{R}'+\Delta\omega)t}\Omega_{+-}(t),
\label{eqn:rotated:frame}
\end{equation}
where $\Delta\omega$ is the total frequency shift given implicitly by the expression
\begin{equation}
\Delta\omega=\mathrm{Re}\int_{0}^{\infty}dt'e^{i(\Omega_{R}'+\Delta\omega)t'}\Sigma_{+-,+-}(t').
\label{eqn:total:freq:shift}
\end{equation}
In the weak coupling regime, $\Omega_{R}'\gg\Delta\omega$, the frequency shift to the leading order in $\Delta\omega$
\begin{equation}
\Delta\omega\simeq\mathrm{Re}\bigg[\Sigma_{+-,+-}(s=-i\Omega_{R}')\bigg].
\label{eqn:total:freq:shift1}
\end{equation}
Here, the purpose of introducing a rotating frame is to get rid of all oscillating parts from $\Omega'(t)$ and hence obtain an equation for $\Omega'_{+-}(t)$
\begin{align}
\dot{\Omega'}_{+-}(t)=&i\Delta\omega\,\Omega'_{+-}(t)-i\int_{0}^{t}dt'\,\tilde{\Sigma}_{+-,+-}(t-t')\Omega'_{+-}(t')\nonumber\\
&-ie^{i(\Omega_{R}'+\Delta\omega)(t+t')}\int_{0}^{t}dt'\,\Sigma_{+-,-+}^{SR}(t-t')\Omega'_{-+}(t'),
\end{align}
where $\tilde{\Sigma}_{+-,+-}(t)=e^{i(\Omega_{R}'+\Delta\omega)t}\Sigma_{+-,+-}(t)$. Due to presence of the oscillatory exponential in the last two terms in RHS of the above equation, we decompose it into slow-varying and fast-oscillating parts as:
\begin{align}
\dot{\Omega}_{+-}^{'S}(t)&=i\Delta\omega\,\Omega'_{+-}(t)-i\int_{0}^{t}dt'\tilde{\Sigma}_{+-,+-}(t-t')\Omega'_{+-}(t')\\
\dot{\Omega}_{+-}^{'F}(t)&=-ie^{i(\Omega_{R}'+\Delta\omega)(t+t')}\int_{0}^{t}dt'\Sigma_{+-,-+}^{SR}(t-t')\Omega'_{-+}(t'),
\end{align}
where S and F stand for slow and fast, respectively. Carrying out Markov approximation on slow term by replacing: $t'\rightarrow t-t'$ , $\Omega'_{+-}(t-t')\rightarrow\Omega'_{+-}(t)$, and finally extending the upper limit of integration to infinity, we obtain
\begin{equation}
\dot{\Omega}_{+-}^{'S}(t)=-\Gamma\,\Omega'_{+-}(t),
\end{equation}
where
\begin{equation}
\Gamma=\frac{1}{T_{2}}=-\mathrm{Im}\int_{0}^{\infty}dt'e^{i(\Omega'_{R}+\Delta\omega)t'}\Sigma_{+-,+-}(t').
\label{eqn:mar:decay:rate}
\end{equation}
Condition for validity of Markov approximation: $e^{i(\Omega'_{R}+\Delta\omega)t'}\Sigma_{+-,+-}(t')$ decays on a time scale $\tau_{c}\ll T_{2}$ which is the decay time of $\Omega_{+-}^{'S}(t)$ and given by the relation\cite{fick}
\begin{equation}
\int_{0}^{\infty}\bigg(\int_{t}^{\infty}dt'e^{i(\Omega'_{R}+\Delta\omega)t'}\Sigma_{+-,+-}(t')\bigg)dt\ll 1.
\end{equation}
On substituting Eqs.~(\ref{eqn:dp:pm}) and (\ref{eqn:trans:pm}) in the above inequality and for $\Gamma_{c}\gg\Omega_{R}', \Delta\omega$ and $\Delta_c$, it leads to the condition $g/\Gamma_{c}\ll 1$ and similarly for the phonon coupling $\alpha_{P}\Delta\omega^{3}e^{\Delta\omega/\epsilon_{c}}(2n_{B}(\Delta\omega)+1)/\epsilon_{c}\ll 1$. Similarly, for the fast term
\begin{equation}
\dot{\Omega}_{+-}^{'F}(t)=-i\,e^{2i(\Omega_{R}'+\Delta\omega)t}\Omega'_{-+}(t)\int_{0}^{\infty}dt'\,\Sigma_{+-,-+}^{SR}(t'),
\label{eqn:mar:decay:rate:1}
\end{equation}
and the condition for validity of Markov approximation:
\begin{equation}
\int_{0}^{\infty}\bigg(\int_{t}^{\infty}dt'\Sigma_{+-,-+}^{SR}(t')\bigg)dt\ll 1,
\end{equation}
which leads to a similar condition $g/\Gamma_{c}\ll 1$. For large $\Omega_{R}'+\Delta\omega$ and due to presence of the highly oscillatory exponential $e^{i(\Omega_{R}'+\Delta\omega)t}$, the effects of terms $\Omega'_{-+}$ will eventually average out to the smaller values compared to $\Omega'_{+-}$. Therefore, in secular approximation when
\begin{equation}
\left|\int_{0}^{\infty}dt'\Sigma_{+-,-+}^{SR}(t')\right|\ll 2(\Omega_{R}'+\Delta\omega),
\label{eqn:secular:1}
\end{equation}
we neglect the fast oscillating terms in Markovian approximation since this term will oscillate fast and average out to a smaller value compared to the slow term. On substituting for $\Sigma_{+-,-+}^{SR}(s)$ from Eq.~(\ref{eqn:trans:mp}) and for $\Gamma_{c}\gg\Delta_{c}$, $\Omega_{R}'\gg\Delta\omega$, we get an explicit condition for the validity of secular approximation as: $g^{2}/(\Omega_{R}'\Gamma_{c})\ll 1$. In the similar manner, we can also neglect the contribution form $\Omega_{++}(t)$ and $\Omega_{--}(t)$ from Eq.~(\ref{eqn:full:eom}). Going back to lab frame and within secular approximation, we obtain the expression in Laplace transform:
\begin{equation}
\Omega_{+-}(s=-i\Delta_{0})\simeq\frac{\Omega_{+-}(0)}{-i(\Delta_{0}-\Omega_{R}'-\Delta\omega)+\Gamma},
\label{eqn:fully:mar:eqn}
 \end{equation}
with initial condition expressed in terms of reduced self-energy matrix elements given by Eqs.~(\ref{eqn:trans:pp}) and (\ref{eqn:trans:mm}), see Appendix~\ref{sec:initial-condition}:
\begin{equation}
\Omega_{+-}(0)=\frac{-{\bf c}^{2}\Sigma_{++,--}^{SR}(s=0)}{\Sigma_{++,++}^{SR}(s=0)-\Sigma_{++,--}^{SR}(s=0)}.
\label{eqn:initial:condition}
\end{equation}
The irrelevant part matrix elements, due to photon $G_{+-,+-}^{R}$ and phonon $G_{+-,+-}^{P}$, both are identically zero under a Markov approximation. On substituting for $\Omega_{+-}(s=-i\Delta_{0})$ from Eq.~(\ref{eqn:fully:mar:eqn}) in the expression (\ref{eqn:one-peak}), we obtain an expression for one-peak Markovian spectrum
\begin{equation}
S_{m}(\Delta_{0})\simeq\frac{X\Gamma}{(\Delta_{0}-\Omega_{R}'-\Delta\omega)^{2}+\Gamma^{2}},
\end{equation}
which is a Lorentzian line centered at $\Delta_{0}=\Omega_{R}'+\Delta\omega$ with a width given by $\Gamma$, frequency-shift $\Delta\omega$ and decay $\Gamma$ are given by Eqs.~(\ref{eqn:total:freq:shift1}) and (\ref{eqn:mar:decay:rate}), respectively, and the pre-factor is given by
\begin{equation}
X=\frac{\bar{I}^{2}{\bf c}^{4}\left(\frac{(\Omega_{R}-\Delta)^{2}}{\Gamma_{c}^{2}+(\Omega_{R}^{'2}+\Delta_{c})^{2}}\right)}{\frac{(\Omega_{R}+\Delta)^{2}}{\Gamma_{c}^{2}+(\Omega_{R}^{'2}-\Delta_{c})^{2}}+\frac{(\Omega_{R}-\Delta)^{2}}{\Gamma_{c}^{2}+(\Omega_{R}^{'2}+\Delta_{c})^{2}}}.
\label{eqn:prefactor}
\end{equation}
Here, we have substituted the expression for the self-energy matrix elements given by Eqs.~(\ref{eqn:trans:pp}) and (\ref{eqn:trans:mm}). Expression for the lineshape in Eq.~(\ref{eqn:fully:mar:eqn}) is Markovian with respect to both photon and phonon interactions. However, we are interested in the non-Markovian regime with respect to phonon coupling and an equation for $\Omega_{+-}(t)$ always contains an extra small term $G_{+-,+-}^{P}(t)$ due to non-Markovian interaction, known as irrelevant part matrix element. Assuming that the irrelevant part is associated with a smallness in the present problem and in order to get the further insight, we will estimate the typical size of its contribution and find a regime where non-Markovian correction is dominant compared to its irrelevant part contribution. Equation for $\Omega_{+-}(t)$ in Laplace domain with its irrelevant part matrix element can be written as,
\begin{widetext}
\begin{equation}
\Omega_{+-}(\Delta_{0})=\frac{1}{-i(\Delta_{0}-\Omega_{R}')+i\Sigma_{+-,+-}^{dP}(\Delta_{0})}\bigg[1+G_{+-,+-}^{P}(\Delta_{0})\bigg]\Omega_{+-}(0).
\label{eqn:laplace:irrelevant:part}
\end{equation}
The smallness of irrelevant part compared to non-Markovian self-energy due to phonon interaction can be justified by the inequality given by
\begin{equation}
\bigg|G_{+-,+-}^{P}(\Delta_{0})\bigg|\ll 1,
\end{equation}
and we only keep the contribution from self-energy matrix element. Furthermore, expanding Eq.~(\ref{eqn:laplace:irrelevant:part}) in the powers of self-energy and ignoring the higher order terms, we have
\begin{equation}
\Omega_{+-}(\Delta_{0})\simeq\frac{1}{-i(\Delta_{0}-\Omega_{R}')}
\bigg[1+\frac{\Sigma_{+-,+-}^{dP}(\Delta_{0})}{\Delta_{0}+\Omega_{R}'}+G_{+-,+-}^{P}(\Delta_{0})\bigg]\Omega_{+-}(0),
\end{equation}
\end{widetext}
and assuming that irrelevant part gives rise to a small contribution and comparing it with self-energy contribution leads to the following inequality,
\begin{equation}
\bigg|\frac{\Sigma_{+-,+-}^{dP}(\Delta_{0})}{\Delta_{0}+\Omega_{R}'}\bigg|\gg\bigg|G_{+-,+-}^{P}(\Delta_{0})\bigg|.
\end{equation}
The one-peak spectrum is centered around $\Delta_{0}\sim\Omega_{R}'$, with a width mostly dominated by Markovian decay rate $\Gamma$, estimating the size of above inequality around $\Delta_{0}\sim\Omega_{R}'+\Gamma$, and it is found that irrelevant part matrix element is always suppressed by a small parameter, $\Gamma/\Omega_{R}'\ll 1$, compared to self-energy matrix element contribution, also see Appendix~\ref{sec:irrelevant-part}. Following the above discussion, we neglect the irrelevant part from the equation for $\Omega_{+-}(t)$ and after going back to lab frame we have an equation written in Laplace domain,
\begin{widetext}
\begin{equation}
\Omega_{+-}(\Delta_{0})=\frac{\Omega_{+-}(0)}{-i[\Delta_{0}-\Omega_{R}'-\Delta\omega_{R}-\Delta\omega_{P}(\Delta_{0})]+\Gamma_{R}+\Gamma_{P}(\Delta_{0})}.
\label{eqn:non:mar:phonon}
\end{equation}
\end{widetext}
Above expression is Markovian with respect to photon coupling but non-Markovian in terms of phonon interaction, where Markovian frequency shift ($\Delta\omega_{R}$) and decay rate ($\Gamma_{R}$) are given by
\begin{align}
\Delta\omega_{R}&\simeq{\rm Re}[\Sigma_{+-,+-}^{dR}(s)+\Sigma_{+-,+-}^{SR}(s)]_{s=-i\Omega_{R}'},\\
\Gamma_{R}&=-{\rm Im}[\Sigma_{+-,+-}^{dR}(s)+\Sigma_{+-,+-}^{SR}(s)]_{s=-i(\Omega_{R}'+\Delta\omega)}.
\end{align}
Similarly, non-Markovian frequency-dependent shift ($\Delta\omega_{P}(\Delta_{0})$) is expressed as
\begin{equation}
\Delta\omega_{P}(\Delta_{0})={\rm Re}[\Sigma_{+-,+-}^{dP}(s)]_{s=-i\Delta_{0}}
\end{equation}
and dephasing ($\Gamma_{P}(\Delta_{0})$) is given by
\begin{equation}
\Gamma_{P}(\Delta_{0})=-{\rm Im}[\Sigma_{+-,+-}^{dP}(s)]_{s=-i\Delta_{0}}.
\end{equation}
\begin{widetext}
On substituting for $\Omega_{+-}(s=-i\Delta_{0})$ from Eq.~(\ref{eqn:non:mar:phonon}) in the expression (\ref{eqn:one-peak}), we obtain an expression for the one-peak non-Markovian spectrum
\begin{equation} S_{nm}(\Delta_{0})=\frac{X[\Gamma_{R}+\Gamma_{P}(\Delta_{0})]}{[\Delta_{0}-\Omega_{R}'-\Delta\omega_{R}-\Delta\omega_{P}(\Delta_{0})]^{2}+[\Gamma_{R}+\Gamma_{P}(\Delta_{0})]^{2}},
\end{equation}
\end{widetext}
where pre-factor $X$ is given by Eq.~(\ref{eqn:prefactor}). It should be noted here that we have not performed Born-Markov approximation in terms of phonon interaction. In Markovian regime frequency shift $\Delta\omega_{P}(\Delta_{0})$ and dephasing $\Gamma_{P}(\Delta_{0})$ are replaced by their $\Delta_{0}=\Omega_{R}'+\Delta\omega$ frequency parts and give rise to an exponential decay and hence to a Lorentzian line centered at $\Delta_{0}=\Omega_{R}'+\Delta\omega$ with a width given by $\Gamma$. Whereas in the non-Markovian regime frequency-shift and dephasing due to phonon interaction are frequency dependent and lead to a non-Markovian (non-exponential) decay giving rise to non-Lorentzian features in the lineshape. We apply above obtained theoretical results to InAs/GaAs QDs, and use the cavity and phonon parameters given in Refs.~\onlinecite{qd1,qd2,ulrich11prl,weiler12,brunner09,krummheuer02}.

\section{\label{sec:results}Results and discussion}
In this section, we plot and analyze the results obtained in the previous sections for different parameter regimes. Typical phonon parameters for GaAs are obtained from Refs.~\onlinecite{weiler12,qd1,qd2}: phonon cut-off $\omega_{c}=1\,{\rm meV}$ and coupling $\alpha_{P}=2.08\times 10^{-7}\,\mu {\rm eV}^{-2}$. For the laser and cavity, we choose following parameters\cite{hughes12prb,ulrich11prl} $\Omega=500\,\mu {\rm eV}$, $g=50\,\mu {\rm eV}$, $\Gamma_{c}=2\,{\rm meV}$. We have varied the other parameters in the plots and explained along with the figures.

\subsection{\label{sec:dephasing}Temperature dependent frequency-shift and dephasing}
In Fig.~\ref{fig:self}, we plot the frequency-shift and dephasing as a function of probe-detuning for different values of phonon bath temperatures. We observe that the effect is strongly temperature-dependent. It increases linearly in the high temperature limit and vanishes for small temperatures. In the present parameter regime for a typical GaAs QD, the dominant contribution from phonon interaction is mainly due to the frequency-shift, see Fig.~\ref{fig:self}. Here, we have assumed that the dot and cavity are resonant and set $\Delta=\Delta_{c}=500\,\mu {\rm eV}$.

\begin{widetext}
\begin{center}
\begin{figure}[ht!]
\centering
\begin{tabular}{cc}
\includegraphics[width=0.5\textwidth]{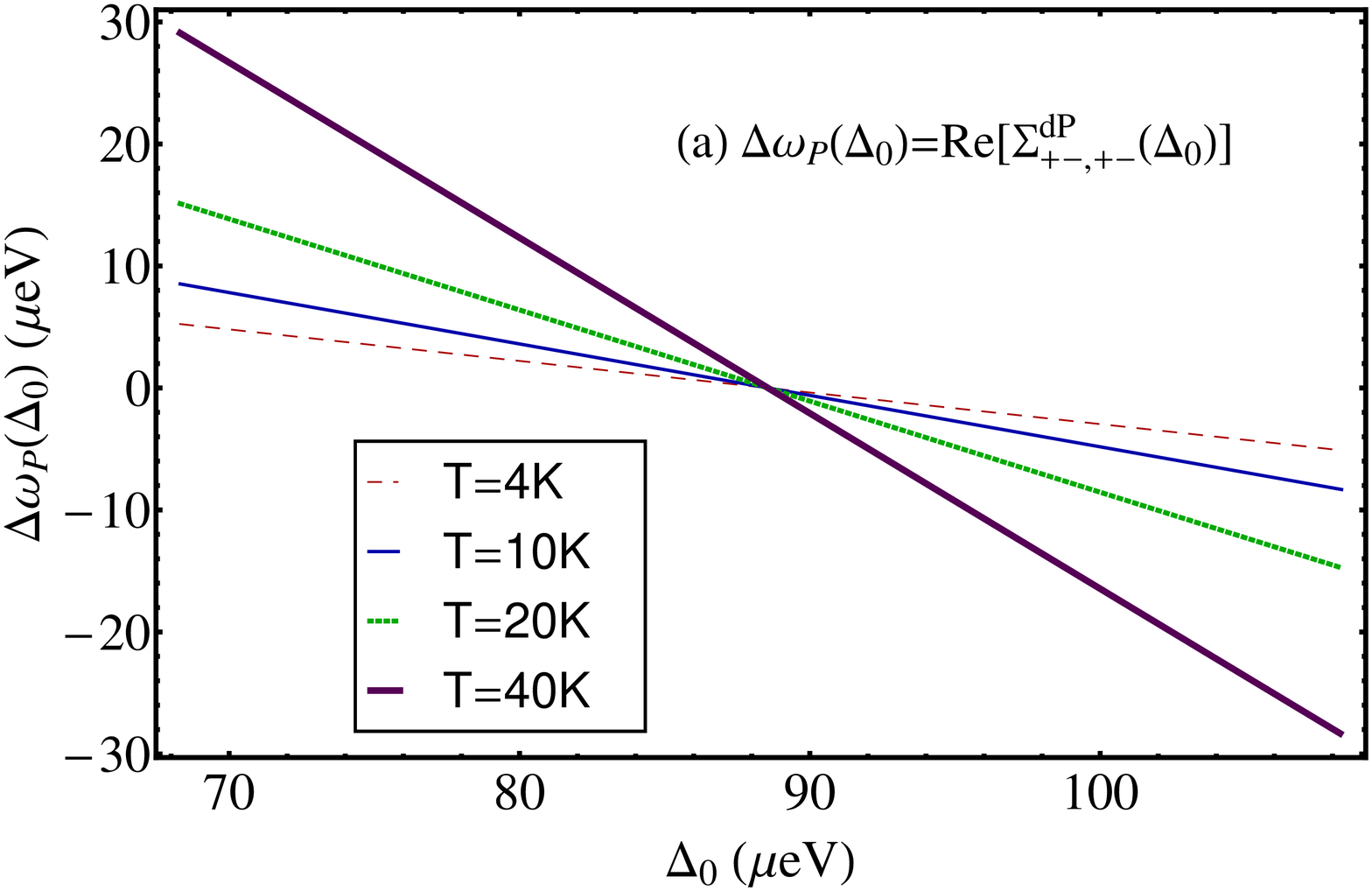} & \includegraphics[width=0.4945\textwidth]{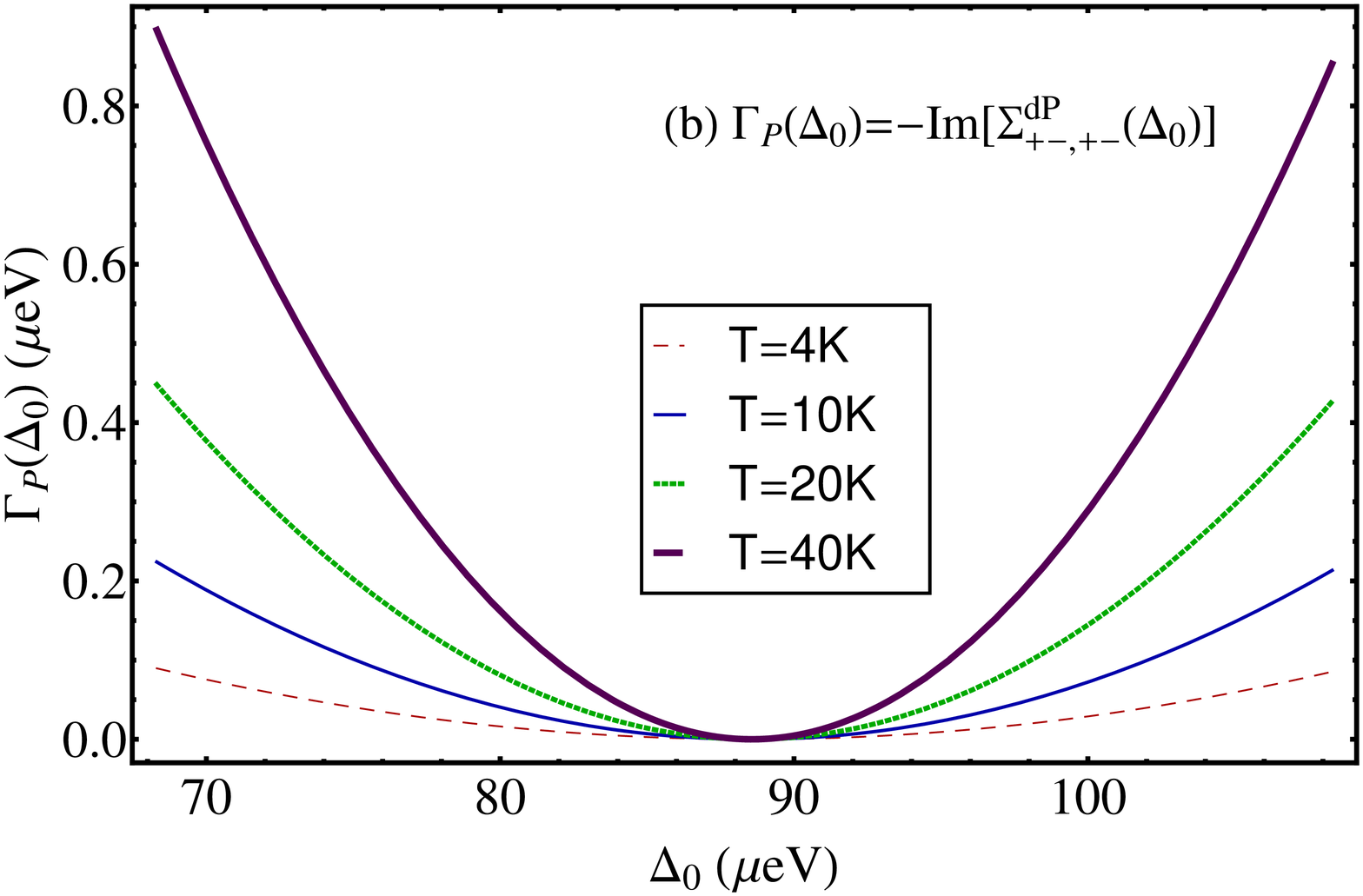}
\end{tabular}
\caption{(Color online) Figures show the frequency-dependent (a) frequency-shift, and (b) dephasing given by the real and imaginary parts of reduced self-energy matrix element, respectively, for the different phonon bath temperatures varying from  $T=4\,{\rm K}$ (flatter one) to $T=40\,{\rm K}$ (steeper one). The frequency-shift and dephasing become steeper in the high temperature limit and vanish for the small temperatures. The cavity-laser and dot-laser detunings are fixed to a value $\Delta=\Delta_{c}=500\mu {\rm eV}$ i.e. dot is resonance with cavity, $\omega_{ab}=\omega_{c}$. Unlike Markovian solution, where the shift and decay are constants and do not change rapidly due a flat cavity band-width, the frequency-shift and dephasing in the non-Markovian regime are strongly probe-dependent bringing a rapid change in the phonon density of states and introducing non-Lorentzian features in the lineshape.}
\label{fig:self}
\end{figure}
\end{center}
\end{widetext}
\subsection{\label{sec:spectra}Temperature dependent one-peak fluorescence spectra}
In Fig.~\ref{fig:abs}, we also plot the associated one-peak spectra for different phonon bath temperatures for the fixed detunings and analyze the effect of frequency-shift and dephasing on the lineshape. We observe a distinct narrowing and asymmetry in the side-peaks with increasing temperature mainly due to the frequency-shift; as frequency-shift changes with increasing temperature giving rise to non-Lorentzian features in the lineshape. This behavior is not observed in the Markovian lineshape as both frequency-shift and dephasing are constants in this case. 

In the dressed-state basis, levels involved in the transitions of interest (Stokes line) are coupled asymmetrically with the phonon modes, and frequency-shift due to phonons pulls these energy levels away from the resonance bringing an additional shift. This extra shift, due to phonons, increases the level separation and reduces the number of channels for the radiative decay. In other words, some of the photons are used to compensate for this additional shift which leads in less photons coming out to the outer world or seen by the detector, eventually causing a narrowing in the sideband. Moreover, the frequency-shift appears to be an odd function with respect to probe detuning (Fig.~\ref{fig:self}) which leads to different probability of emitting and absorbing the phonons on the either side of probe detuning.
\begin{widetext}
\begin{center}
\begin{figure}[H]
\centering
\begin{tabular}{cc}
\includegraphics[width=0.5\textwidth]{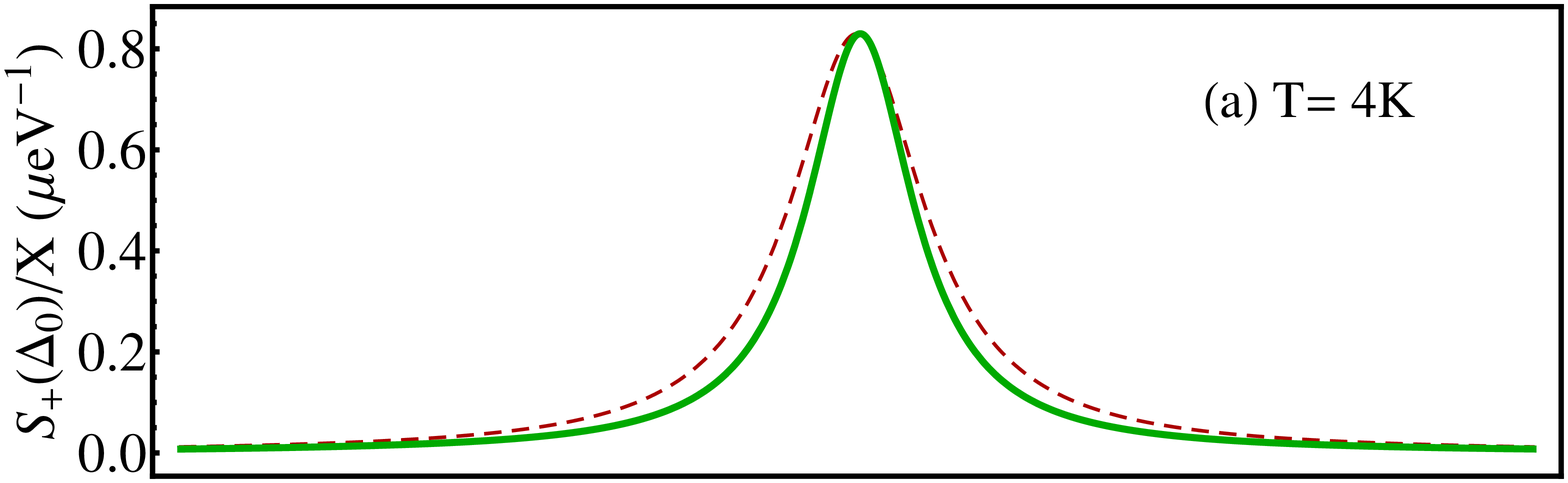} & \includegraphics[width=0.5\textwidth]{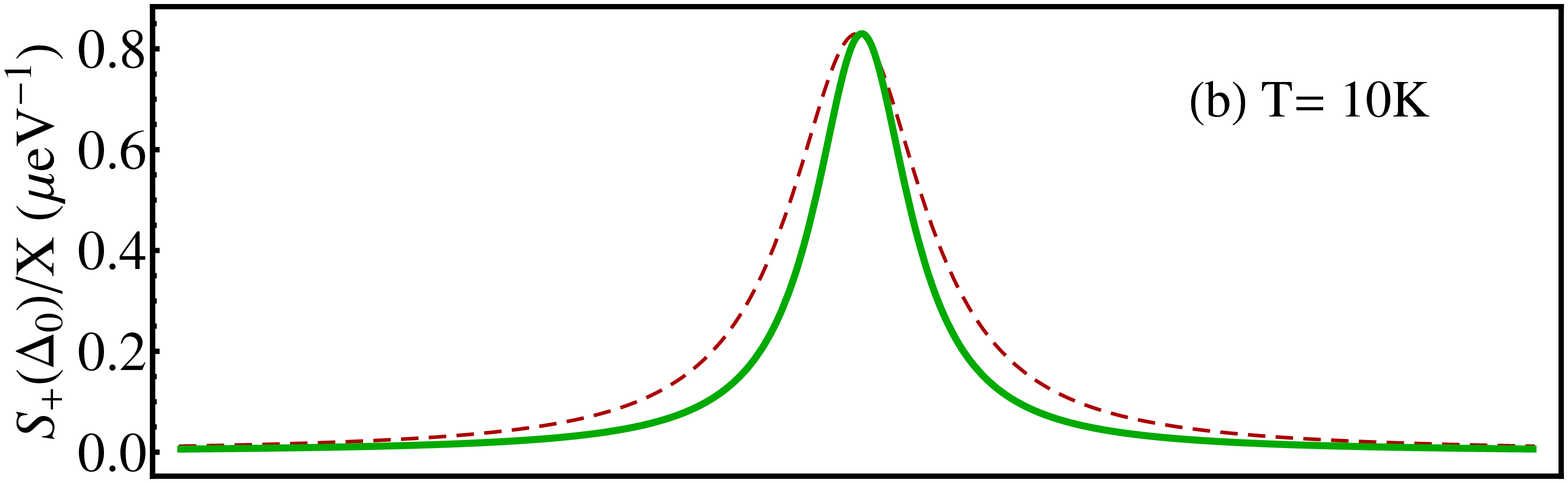}\\
\includegraphics[width=0.5\textwidth]{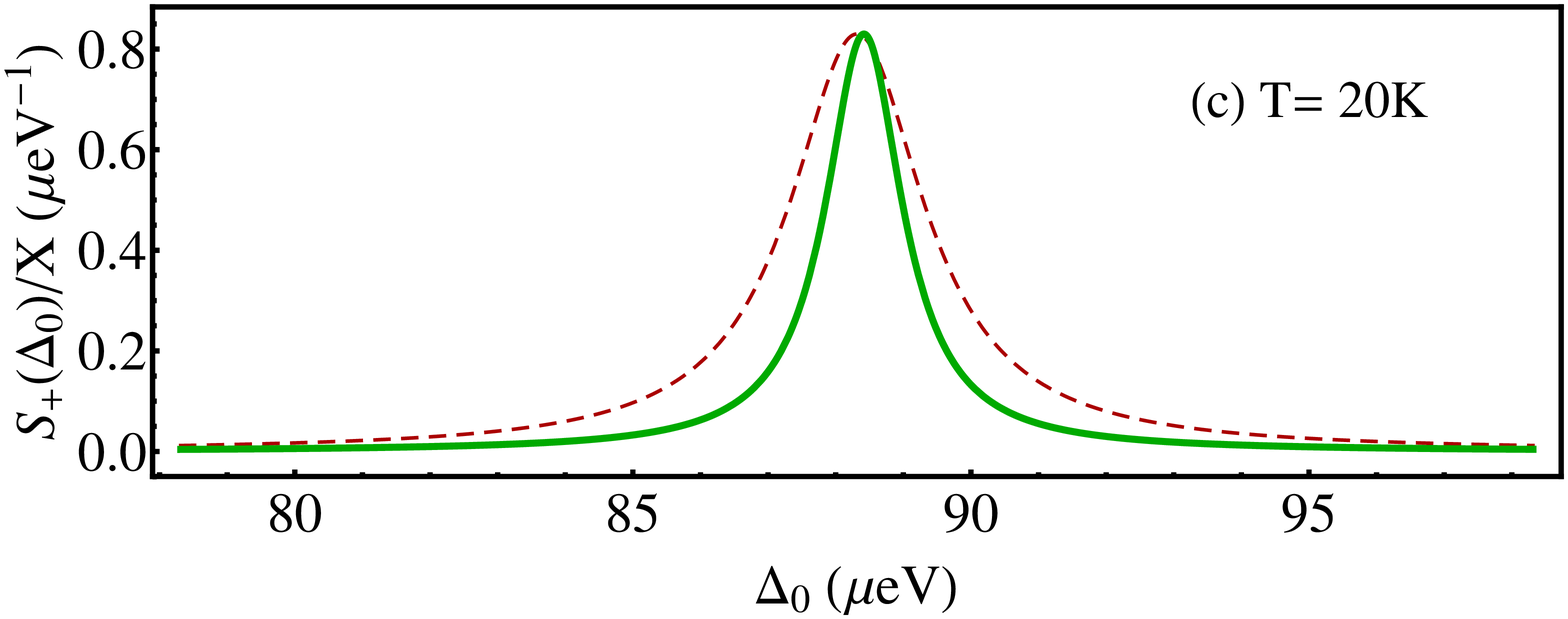} & \includegraphics[width=0.5\textwidth]{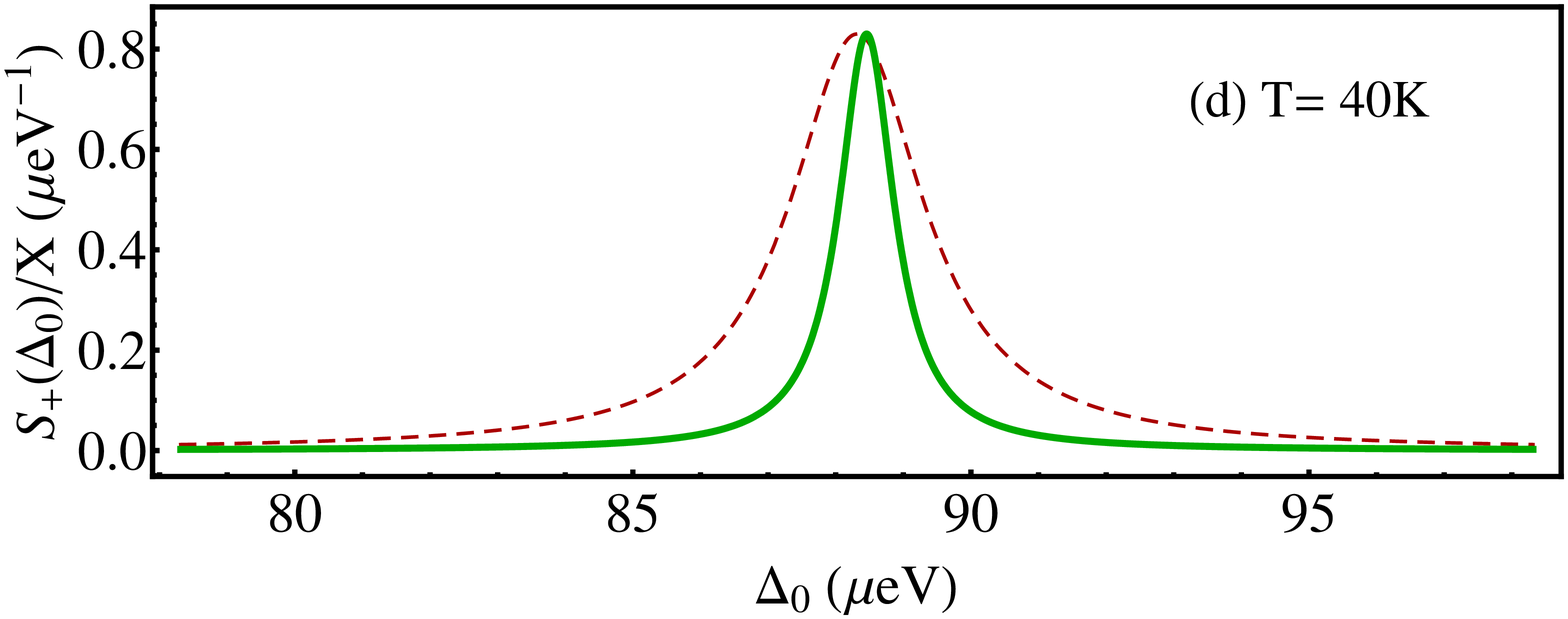}
\end{tabular}
\caption{ (Color online) Figures show a comparison between one-peak fluorescence spectra for the different phonon bath temperatures, obtained within Markovian (dashed lines) and non-Markovian (solid lines) regimes. The cavity-laser and dot-laser detunings are fixed to a value $\Delta=\Delta_{c}=500\mu eV$, and other parameters are same as in Fig.~\ref{fig:self}. Distinct narrowing and asymmetry in the side-peak are observed due to phonon interaction, and are described in our theoretical model.}
\label{fig:abs}
\end{figure}
\end{center}
\end{widetext}

We also observe that the Lamb shift and additional broadening, due to phonon interaction, vanish for the case of resonant pump laser. 
This can be explained in the \emph{dressed-state} representation, where both energy-levels of interest, correspond to one of the transitions (Stokes line), are coupled asymmetrically to the phonons. For the zero detuning they become degenerate leading to the vanishing frequency-shift and dephasing. We have also observed a little shifts in the peak positions due to strong dependence of frequency-shift on the probe-detuning.

\section{\label{sec:conclusions}Conclusion}
We have discussed the dynamics of a driven cavity-QED system coupled to a 3d acoustic phonon reservoir with non-Markovian pure dephasing mechanism. We have observed highly modified non-Lorentzian features in the associated spectrum due to phonon interaction, and for solid-state systems it can only be described within non-Markovian regime. We have shown that quantum regression theorem is not valid in this case because of a non-zero irrelevant part, and also the full density matrix is \emph{not} separable for all times. The system correlation is solved without using the quantum regression theorem beyond usual Born-Markov approximation. We have obtained the analytical formulas for both Markovian and non-Markovian one-peak lineshapes in terms of the model parameters. Both expressions look like Lorentzian lines centered around $\Delta_{0}\simeq\Omega_{R}'$, but this analogy is not correct as shift and dephasing are also probe-dependent in case of non-Markovian lineshape. We have investigated the one-peak spectrum of the Mollow-triplet for different values of phonon bath temperature for the resonant dot and cavity. We have found vanishing shift and dephasing when laser is resonant with the dot, because levels are equally and asymmetrically coupled to phonon modes rendering them degenerate for a resonant pump laser. We have derived an exact form of Najakjima-Zwanzig generalized master equation and showed that Markov and non-Markov solutions give significantly different results. We show that non-Markovian contribution is significantly large and can be clearly seen in the spectrum. A distinct narrowing of the sideband has been reported which is contrary to the recent results leading to broadening in the presence of phonons. We have shown that the frequency-dependent shift has strong temperature dependence which causes different features like narrowing and asymmetry in the lineshape. This procedure can also be used to systematically account for features in optical spectra of a general multi-level system due to genuine non-Markovian dynamics.

\section*{Acknowledgments}
AK acknowledge financial support from the Icelandic Research Fund RANNIS, CIFAR, Canada, the National Key Research and Development Program of China (Grant No. 2016YFA0301200), and the NFSC grants (No 11574025 and No. U1530401). AK thanks Bill Coish, Sigurdur I. Erlingsson, Stefano Chesi, Li-jing Jin, and Tilen Cadez for the useful discussions and feedback.
\appendix

\section{\label{sec:detector response}Detector response function}
The detector response function in Eq.~(\ref{eqn:flou:spec:11})
\begin{equation}
\bar{I}=\int_{0}^{\infty}d\tau\,\sum_{k}g_{k}g_{k}^{D}e^{-i\omega_{k}\tau},
\end{equation}
where $g_{k}^{D}=|\mathcal{\wp}_{\alpha\beta}|\varepsilon_{k}$. Applying the continuum of modes after replacing the sum via integral $\sum_{k}g_{k}^{2}\to\int_{0}^{\infty}D(\epsilon)|g(\epsilon)|^{2}d\epsilon$ and using the well-known formula
\begin{equation}
\frac{1}{X\pm i0^{+}}=\mathcal{P}\left(\frac{1}{X}\right)\mp i\pi\delta(X),
\end{equation}
($\mathcal{P}$ indicates the principal part) one can write the square of detector response function in terms of principal part and delta function as
\begin{align}
\bar{I}^{2}=&\int_{0}^{\infty}d\epsilon\,D_{c}(\epsilon)|g(\epsilon)|^{2}\bigg[-i\mathcal{P}\left(\frac{1}{\epsilon}\right)+\pi\delta(\epsilon)\bigg]\nonumber\\
&\times\int_{0}^{\infty}d\nu\,D(\nu)|g^{D}(\nu)|^{2}\bigg[i\mathcal{P}\left(\frac{1}{\nu}\right)+\pi\delta(\nu)\bigg],
\end{align}
where $D_{c}(\epsilon)$ and $D(\nu)$ are photonic density of states of cavity and open space, respectively. The photonic density of states of the cavity is described by a Lorentzian density of states given by Eq.~(\ref{eqn:dos:cavity}).

\section{\label{sec:schrieffer-wolff}Schreiffer-Wolff Transformation}
Hamiltonian $H'$ resulting from the polaron transformation, is 
\begin{align}
H'=&\frac{\Omega_{R}'}{2}\sigma_{3}+\sum_{k}\Delta_{k}a_{k}^{\dag}a_{k}+\sum_{q}\lambda_{q}b_{q}^{\dag}b_{q}\nonumber\\
&+{\bf c.s}\,\sigma_{3}\,\sum_{k}g_{k}(a_{k}+a_{k}^{\dag})\nonumber\\
&+\frac{({\bf c}^{2}-{\bf s}^{2})}{2}\sigma_{3}\,\sum_{q}\lambda_{q}(b_{q}+b_{q}^{\dag})\nonumber\\
&+\sum_{k}g_{k}\left[({\bf c}^{2}a_{k}-{\bf s}^{2}a_{k}^{\dagger})\sigma_{+-}+h.c.\right]\nonumber\\
&-{\bf c.s}\,\sum_{q}\lambda_{q}(\sigma_{+-}+\sigma_{-+})(b_{q}+b_{q}^{\dagger})\nonumber\\
&-\sum_{q}\frac{\lambda_{q}^{2}}{4\omega_{q}}+{\bf c.s}\,\sum_{q}\frac{\lambda_{q} ^{2}}{\omega_{q}}(\sigma_{+-}+\sigma_{-+}).
\end{align}
Last two terms in above Hamiltonian commute with rest of the Hamiltonian and can be igonred within a secular approximation for large $\Omega_{R}'$. We get rid of the energy exchange process due to phonon interaction using leading-order Schrieffer-Wolff transformation\cite{swt} and start from the full Hamiltonian:
\begin{align}
H'&=H_{1}+V_{2}\\
H_{1}&=H_{S}+H_{R}+H_{P}+H_{dR}+H_{dP}+H_{SR},
\end{align}
where individual terms are defined as
\begin{align}
H_{S}&=\frac{\Omega'_{R}}{2}\sigma_{3},\\
H_{R}&=\sum_{k}\Delta_{k}a_{k}^{\dag}a_{k},\\
H_{P}&=\sum_{q}\lambda_{q}b_{q}^{\dag}b_{q},\\
H_{dR}&={\bf c.s}\,\sigma_{3}\,\sum_{k}g_{k}(a_{k}+a_{k}^{\dag}),\\
H_{dP}&=\frac{({\bf c}^{2}-{\bf s}^{2})}{2}\,\sigma_{3}\,\sum_{q}\lambda_{q}(b_{q}+b_{q}^{\dag}),\\
H_{SR}&=\sum_{k}g_{k}\left[({\bf c}^{2}a_{k}-{\bf s}^{2}a_{k}^{\dagger})\sigma_{+-}+h.c.\right],\\
V_{2}&=-{\bf c.s}\,\sum_{q}\lambda_{q}(\sigma_{+-}+\sigma_{-+})(b_{q}+b_{q}^{\dagger}).
\end{align}
We apply a transformation, $\bar{H}=e^{A}He^{-A}$, generated by an anti-hermitian operator $A=-A^{\dag}$ to eliminate the transition terms to the first order. Using Baker-Campbell-Hausdorff formula and expanding $\bar{H}$ in the powers of $A$, we obtain
\begin{equation}
\bar{H}=H_{1}+V_{2}+[A,H_{1}]+[A,V_{2}]+\frac12[A,[A,H]]+...
\end{equation}
In order to get rid of transition term $V_{2}$ to leading order, we set $V_{2}=-[A,H_{1}]$, where $A$ can be written as
\begin{equation}
A=\frac{1}{L_{1}}V_{2},
\end{equation}
and $L_{1}\mathcal{O}=[H_{1},\mathcal{O}]$. Here $A$ is of order of transition term $V_{2}$. Substituting for $A$, we obtain Hamiltonian up to the second or higher order in $V_{2}$
\begin{equation}
\bar{H}=H_{1}+\frac12[A,V_{2}]+... 
\end{equation}
Using the definitions of $H_{1}$ and $V_{2}$, we obtain the expression for A as:
\begin{equation}
A=-{\rm c.s}\sum\limits_{q}\frac{\lambda_{q}}{\Omega_{R}}(b_{q}+b_{q}^{\dag})(\sigma_{+-}-\sigma_{-+}).
\end{equation}
Therefore, the transformed Hamiltonian to the first order in the transition terms due to phonon can be well approximated and written as free and perturbed parts as
\begin{equation}
\bar{H}\simeq H_{0}+H_{V},
\end{equation} 
where the free part is
\begin{align}
H_{0}&=H_{S}+H_{R}+H_{P},\\
H_{S}&=\frac{\Omega'_{R}}{2}\sigma_{3},\\
H_{R}&=\sum_{k}\Delta_{k}a_{k}^{\dag}a_{k},\\
H_{P}&=\sum_{q}\omega_{q}b_{q}^{\dag}b_{q}
\end{align}
and perturbed parts is given by
\begin{align}
H_{V}&=H_{dR}+H_{SR}+H_{dP},\\
H_{dR}&={\bf c.s}\,\sigma_{3}\,\sum_{k}g_{k}(a_{k}+a_{k}^{\dag}),\\
H_{dP}&=\frac{{\bf c}^{2}-{\bf s}^{2}}{2}\,\sigma_{3}\,\sum_{q}\lambda_{q}(b_{q}+b_{q}^{\dag}),\\
H_{SR}&=\sum_{k}g_{k}\left[({\bf c}^{2} a_{k}-{\bf s}^{2} a_{k}^{\dagger})\sigma_{+-}+h.c.\right],
\end{align}
which are Eqs.~(\ref{eqn:polaron:hamiltonian})-(\ref{eqn:trans:rad}) in the main text.

\section{\label{sec:irrelevant-part}Irrelevant part matrix element}
The irrelevant part is given by Eqn.\ (\ref{eqn:irr:part}) can be written within Born approximation after transforming to Laplace domain
\begin{equation}
\Phi(s)\simeq-i{\rm Tr_{R}Tr_{P}}L_{V}\,\frac{1}{s+iL_{0}}\,Q\Omega(0).
\end{equation}
In particular, we want the matrix element, $\Phi_{+-,+-}(s)$ due to phonon interaction, which can be simplified and written as
\begin{equation}
\Phi_{+-,+-}(s)=-i{\rm Tr_{P}}\,L_{Y}^{+}\,\frac{1}{s+i(\Omega_{R}'+L_{P})}\,[Q\Omega(0)]_{+-},
\end{equation}
where irrelevant part of the stationary density matrix can be found\cite{swan81} using GME discussed in Sec.~\ref{sec:gme}
\begin{equation}
[Q\Omega(0)]_{+-}=-i\lim_{s\rightarrow 0}\frac{s}{s+i(\Omega_{R}'+L_{P})}L_{Y}^{+}\rho_{P}(t_0)\,[\rho_{S}(s)\sigma_{ab}]_{+-}.
\label{eqn:irrelevant part matrix element}
\end{equation}
The propagators in Eqn.\ (\ref{eqn:irrelevant part matrix element}) has no poles at $s=0^{+}$, where $0^{+}$ is a positive infinitesimal. After performing the limit in the above expression, we substitute for $[Q\Omega(0)]_{+-}$ in the expression for $\Phi_{+-,+-}(s)$, to obtain and expression for the irrelevant term as
\begin{widetext}
\begin{equation}
\Phi_{+-,+-}(s)=-{\rm Tr_{P}}\,L_{Y}^{+}\,\frac{1}{s+i(\Omega_{R}'+L_{P})}\frac{1}{0^{+}+i(\Omega_{R}'+L_{P})}L_{Y}^{+}\rho_{P}(t_0)\Omega_{+-}(0),
\end{equation}
where $\Omega_{+-}(0)=[\bar{\rho}_{S}\sigma_{ab}]_{+-}$, see Eqs.~(\ref{eqn:initial:condition:matrix}) and (\ref{eqn:initial:condition:matrix:element}). Above expression can be written in terms of irrelevant part matrix element in the problem as 
\begin{align}
\Phi_{+-,+-}(s)&=G_{+-,+-}^{P}(s)\Omega_{+-}(0),\\
G_{+-,+-}^{P}(s)&=-{\rm Tr_{P}}\,L_{Y}^{+}\,\frac{1}{s+i(\Omega_{R}'+L_{P})}\frac{1}{0^{+}+i(\Omega_{R}'+L_{P})}L_{Y}^{+}\rho_{P}(t_0).
\end{align}
\end{widetext}
Solving above expressions and applying the continuum of modes, we found that irrelevant part matrix element will be suppressed by $1/\Omega_{R}'$ compared to the self-energy matrix element due to phonon interaction. In this limit, contribution from irrelevant part can be neglected compared to the contribution from the non-Markovian self-energy.

\section{\label{sec:self-energy-calculations} Self-energy calculations}
The reduced self-energy superoperator $\Sigma_{S}(t)$ in Eq.~(\ref{eqn:reduced:self:energy}) can be transformed in to Laplace domain and using $L_{V}=L_{dR}+L_{SR}+L_{dP}$, we obtain
\begin{align}
\Sigma_{S}(s)=&-i{\rm Tr_{R}Tr_{P}}(L_{dR}+L_{SR}+L_{dP})\frac{1}{s+iL}\nonumber\\
&\times(L_{dR}+L_{SR}+L_{dP})\rho_{R}(t_{0})\rho_{P}(t_{0}),
\label{eqn:decomposed:self:energy}
\end{align}
dropping $Q$ from the exponential in Eq.~(\ref{eqn:reduced:self:energy}) will not affect the final expression \cite{loss05}. Cross terms in above expression will vanish because $L_{dR}$ and $L_{SR}$ act on an operator in the radiation mode Hilbert space whereas $L_{dP}$ act on phonon Hilbert space and will not contribute in the final trace. Moreover, the cross terms between $L_{dR}$ and $L_{SR}$ will give rise to off-block diagonal matrix elements in self-energy matrix and will be neglected within secular approximation, see Sec.~\ref{sec:secular:approximation} in main text. The self-energy superoperator in Eq.~(\ref{eqn:decomposed:self:energy}) can be decomposed into three different parts as:
\begin{equation}
\Sigma_{S}(s)=\Sigma_{S}^{dR}(s)+\Sigma_{S}^{dP}(s)+\Sigma_{S}^{SR}(s),
\end{equation}
where first and second terms give rise to pure dephasing ($T_{2}^{*}$ process) due to radiation and phonon modes, respectively, whereas last term in above expression leads to transition ($T_{1}$ process) due to radiation modes coupling. Free propagator in reduced self-energy expression can be expanded in the powers of interacting Liouvillian $L_{V}$ as \cite{bill04}
\begin{equation}
\frac{1}{s+iL}=\frac{1}{s+iL_{0}}\sum_{k}\bigg(-iL_{V}\frac{1}{s+iL_{0}}\bigg)^{2k},
\end{equation}
because of the form of couplings in present model only even powers $2k$ in above expression will survive in the final trace. In order to find the matrix elements of the self-energy superoperator, we write the superoperators in matrix form in dressed-state basis as
\begin{equation}
[L_{S}]=
\left(
\begin{array}{cccc}
0 & 0 & 0 & 0\\
0 & 0 & 0 & 0\\
0 & 0 & \Omega_{R}' & 0\\
0 & 0 & 0 & -\Omega_{R}'
\end{array}
\right),
\end{equation}
where $[L_{S}]_{\alpha\beta,\gamma\delta}=Tr\lbrace\ket{\beta}\bra{\alpha}S\ket{\gamma}\bra{\delta}\rbrace$ and $\lbrace\alpha,\beta\rbrace\in\lbrace +,-\rbrace$. In the dressed state basis, non-interacting Liouvillian is diagonal and can be inverted to write as $2\times 2$ blocks
\begin{equation}
\bigg[\frac{1}{s+iL_{0}}\bigg]=
\left(
\begin{array}{cc}
G_{\parallel}(s) & 0\\
0 & G_{\perp}(s)
\end{array}
\right),
\end{equation}
where parallel block is
\begin{equation}
[G_{\parallel}(s)]=
\left(
\begin{array}{cc}
\frac{1}{s+i(L_{R}+L_{P})} & 0\\
0 & \frac{1}{s+i(L_{R}+L_{P})}
\end{array}
\right),
\end{equation}
and perpendicular block is given by
\begin{equation}
[G_{\perp}(s)]=
\left(
\begin{array}{cc}
\frac{1}{s+i(\Omega_{R}'+L_{R}+L_{P})} & 0\\
0 & \frac{1}{s+i(-\Omega_{R}'+L_{R}+L_{P})}
\end{array}
\right).
\end{equation}
In the similar fashion, we can find other matrices as well
\begin{equation}
[L_{dR(P)}]=
\left(
\begin{array}{cccc}
L_{X(Y)}^{-} & 0 & 0 & 0\\
0 & L_{X(Y)}^{-} & 0 & 0\\
0 & 0 & L_{X(Y)}^{+} & 0\\
0 & 0 & 0 & -L_{X(Y)}^{+}\\
\end{array}
\right),
\end{equation}
here we have defined new Liouvillian for commutation and anti-commutation relations: $L_{X(Y)}^{\pm}\mathcal{O}=[X_{R(P)},\mathcal{O}]_{\pm}$, where operators $X_{R}$ and $X_{P}$ are given as
\begin{align}
X_{R}&=\frac{\sqrt{\Omega_{R}^{2}-\Delta^{2}}}{2\Omega_{R}}\sum_{k}g_{k}(a_{k}+a_{k}^{\dag}),\\
X_{P}&=\frac{\Delta}{2\Omega_{R}}\sum_{k}\lambda_{q}(b_{q}+b_{q}^{\dag}).
\end{align}
Furthermore, we also find the matrix for superoperator $L_{SR}$ in the dressed-state basis
\begin{equation}
[L_{SR}]=
\left (
\begin{array}{cccc}
0 & 0 & -Z_{r}^{\dag} & Z_{l}\\
0 & 0 & Z_{l}^{\dag} & -Z_{r}\\
-Z_{r} & Z_{l} & 0 & 0\\
Z_{l}^{\dag} & -Z_{r}^{\dag} & 0 & 0
\end{array}
\right ),
\end{equation}
where we have defined the operators for left and right multiplications as:
\begin{align}
Z_{l}\mathcal{O}_{R}&=\sum\limits_{k}g_{k}({\bf c}^{2}a_{k}-{\bf s}^{2}a_{k}^{\dag})\mathcal{O}_{R}\\
Z_{r}\mathcal{O}_{R}&=\mathcal{O}_{R}\sum\limits_{k}g_{k}({\bf c}^{2}a_{k}-{\bf s}^{2}a_{k}^{\dag}),
\end{align}
and $\mathcal{O}_{R}$ is an operator in the radiation mode Hilbert space. Reduced self-energy matrix elements of interest can be calculated according to Eq.~(\ref{eqn:full:eom}) in the main text.
\subsection{\label{sec:photon-self-energy}Self-energy for photon interaction}
We apply a continuum of modes for the cavity density of states given by a Lorentzian spectrum \cite{nori12},
\begin{equation}
D_{c}(\epsilon)|g(\epsilon)|^{2}=\frac{1}{\pi}\frac{g^{2}\Gamma_{c}}{(\epsilon-\Delta_{c})^{2}+\Gamma_{c}^{2}},
\label{eqn:dos:cavity}
\end{equation}
where $\Delta_{c}=\omega_{c}-\omega$ is the detuning of cavity from laser pump frequency, and $\Gamma_{c}$ is cavity bandwidth. Reduced self-energy matrix elements due to radiation mode coupling giving rise to transition are given as
\begin{widetext}
\begin{align}
\Sigma_{+-,++}^{SR}(s)&=\frac{-ig^{2}(\Omega_{R}+\Delta)^{2}}{4\Omega_{R}^{2}}\left(\frac{1}{s+i(\Omega_{R}'-\Delta_{c})+\Gamma_{c}}+\frac{1}{s-i(\Omega_{R}'-\Delta_{c})+\Gamma_{c}}\right),\\
\nonumber\\
\Sigma_{+-,--}^{SR}(s)&=\frac{ig^{2}(\Omega_{R}-\Delta)^{2}}{4\Omega_{R}^{2}}\left(\frac{1}{s+i(\Omega_{R}'+\Delta_{c})+\Gamma_{c}}+\frac{1}{s-i(\Omega_{R}'+\Delta_{c})+\Gamma_{c}}\right),\\
\nonumber\\
\Sigma_{+-,+-}^{SR}(s)&=\frac{-ig^{2}}{4\Omega_{R}^{2}}\bigg(\frac{(\Omega_{R}+\Delta)^{2}}{s+i\Delta_{c}+\Gamma_{c}}+\frac{(\Omega_{R}-\Delta)^{2}}{s-i\Delta_{c}+\Gamma_{c}}\bigg),\\
\nonumber\\
\Sigma_{+-,-+}^{SR}(s)&=\frac{-ig^{2}(\Omega_{R}^{2}-\Delta^{2})}{4\Omega_{R}^{2}}\bigg(\frac{1}{s+i\Delta_{c}+\Gamma_{c}}+\frac{1}{s-i\Delta_{c}+\Gamma_{c}}\bigg),
\end{align}
and self-energy matrix element responsible for pure dephasing due to cavity coupling can be found as
\begin{equation}
\Sigma_{+-,+-}^{dR}(s)=\frac{-ig^{2}(\Omega_{R}^{2}-\Delta^{2})}{2\Omega_{R}^{2}} \left(\frac{1}{s+i(\Omega_{R}'-\Delta_{c})+\Gamma_{c}}+\frac{1}{s+i(\Omega'_{R}+\Delta_{c})+\Gamma_{c}}\right).
\end{equation}
\end{widetext}
For a large band-width cavity, coupling to its radiative modes to system is treated under Markov approximation and above self-energies are replaced by their $s=-i(\Omega_{R}'+\Delta\omega)$ frequency parts, refer main text for details.
\subsection{\label{sec:phonon-self-energy}Self-energy for phonon interaction}
Similarly, we apply a continuum of modes for 3-d acoustic phonons\cite{krummheuer02} with an exponential cut-off at $\epsilon=\epsilon_{c}$ 
\begin{equation}
\sum_{q}\lambda_{q}^{2}\to\alpha_{P} \int_{0}^{\infty}d\epsilon|\epsilon|^{3}e^{-|\epsilon|/\epsilon_{c}},
\end{equation}
we obtain the expression for self-energy in Laplace transform as
\begin{align}
\Sigma_{+-,+-}^{dP}(s)=&\frac{-i\alpha_{P}\Delta^{2}}{2\Omega_{R}^{2}}\int_{0}^{\infty}d\epsilon|\epsilon|^{3}e^{\frac{-|\epsilon|}{\epsilon_{c}}}(2n_{B}(\epsilon)+1)\nonumber\\
&\times\left(\frac{1}{s+i(\Omega_{R}'-\epsilon)}+\frac{1}{s+i(\Omega_{R}'+\epsilon)}\right),
\label{eqn:pdephas:pm}
\end{align}
where $\alpha_{P}=$ is the phonon coupling paramter in the units of $\mathrm{freq.}^{-2}$ and $n_{B}(\epsilon)$ is Bose function. On further simplification, above self-energy matrix element can be decomposed into real and imaginary parts after setting $s=-i\Delta_{0}$, where $\Delta_{0}$ is detuning of the probe from the pump laser frequency, as:
\begin{equation}
\Sigma_{+-,+-}^{dP}(s=-i\Delta_{0})=\Delta\omega_{P}(\Delta_{0})-i\Gamma_{P}(\Delta_{0}),
\label{eqn:complex:self:energy}
\end{equation}
where $\Delta\omega_{P}(\Delta_{0})={\rm Re}[\Sigma_{+-,+-}^{dP}(\Delta_{0})]$ is the frequency-shift and $\Gamma_{P}(\Delta_{0})=-{\rm Im}[\Sigma_{+-,+-}^{dP}(\Delta_{0})]$ is the dephasing due to phonon interaction.
\subsection{\label{sec:exact:self:energy}Self-energy matrix element calculated exactly}
In the previous section, we have computed the self-energy matrix element for phonon interaction to the second order in Born approximation. In this section, we will discuss an equation-of-motion method to find the phonon interaction self-energy for all orders in perturbed Liouvillain due to phonons $L_{Y}^{+}$ beyond Born approximation, and show that exact approach recovers the result obtained within Born approximation. Using the general form of superoperators matrices, the expression for self-energy matrix element due to phonons can be written in Laplace domain as:
\begin{equation}
\Sigma_{+-,+-}^{P}(s)=-i{\rm Tr_{P}}\,L_{Y}^{+}\,\frac{1}{s+i(\Omega_{R}'+L_{P}+L_{Y}^{+})}\,L_{Y}^{+}\,\rho_{P}(t_0),
\end{equation}
also in the time domain, we have
\begin{equation}
\Sigma_{+-,+-}^{P}(t)=-ie^{-i\Omega_{R}'t} \underbrace{{\rm Tr_{P}}\,\big[L_{Y}^{+}\,e^{-i(L_{P}+L_{Y}^{+})}\,L_{Y}^{+}\,\rho_{P}(t_0)\big]}_{\mathcal{C}(t)}.
\label{Sigma_phonon_exact}
\end{equation}
On further simplification, one obtains
\begin{align}
\mathcal{C}(t)&=2Tr_{P}\,\biggl[[X_{P},X_{P}(t)]_{+}\rho_{P}(0)\biggr]\nonumber\\
&=2\langle X_{P}X_{P}(t)\rangle+2\langle X_{P}(t)X_{P}\rangle,
\end{align}
where
\begin{equation}
X_{P}(t)=e^{-i(L_{P}+L_{Y}^{+})t}X_{P}(0).
\end{equation}
Above expression gives rise to a differential equation
\begin{equation}
\dot{X}_{P}(t)=-i(L_{P}+L_{Y}^{+})X_{P}(t),
\end{equation}
which can be written as
\begin{equation}
\dot{X}_{P}(t)=-i(L_{Y}^{+}-L_{P})X_{P}(t).
\label{eqn:1}
\end{equation}
Introducing: $\tilde{X}_{P}(t)=e^{-iL_{P}t}X_{P}(t)$, we obtain an equation of motion for $\tilde{X}_{P}(t)$
\begin{equation}
\dot{\tilde{X}}_{P}(t)=-i[ X_{P}^{0}(t),\tilde{X}_{P}(t)]_{+}
\label{eqn:eom in ques}
\end{equation}
where 
\begin{align}
X_{P}^{0}(t)=\frac{\Delta}{2\Omega_{R}}\sum_{q}\lambda_{q}(b_{q}e^{i\omega_{q}t}+b_{q}^{\dag}e^{-i\omega_{q}t}).
\label{eqn:eom with hq}
\end{align}
Solving for one $q$, the solution for $\tilde{X}_{P}(t)$ takes the form
\begin{equation}
\tilde{X}_{P,q}(t)=U_{q}(t)\tilde{X}_{P,q}(0)W_{q}(t)
\end{equation}
where
\begin{align}
\tilde{X}_{P}(t)&=\sum_{q}\tilde{X}_{P,q}(t)\\
\dot{U}_{q}(t)&=-iX_{P,q}^{0}(t)U_{q}(t)\label{eqn:eom uq}\\
\dot{W}_{q}(t)&=-iW_{q}(t)X_{P,q}^{0}(t)\label{eqn:eom uw}.
\end{align}
Using Eqns.\ (\ref{eqn:eom with hq}) and (\ref{eqn:eom uq}), we have (for one $q$)
\begin{equation}
\dot{U}_{q}(t)=-i\lambda_{q}^{'}e^{-iH_{q}t}(b_{q}+b_{q}^{\dag})e^{iH_{q}t}U_{q}(t)
\label{eqn:3}
\end{equation}
where
\begin{align}
\lambda_{q}^{'}&=\frac{\Delta\lambda_{q}}{2\Omega_{R}}\\
H_{q}&=\omega_{q}b_{q}^{\dag}b_{q}.
\end{align}
Introducing
\begin{equation}
\tilde{U}_{q}(t)=e^{iH_{q}t}U_{q}(t)\quad\Rightarrow U_{q}(t)=e^{-iH_{q}t}\tilde{U}_{q}(t)
\end{equation}
and taking the time derivative, one can find an expression
\begin{equation}
\dot{\tilde{U}}_{q}(t)=i[H_{q}-\lambda_{q}^{'}(b_{q}+b_{q}^{\dag})]\tilde{U}_{q}(t).
\end{equation}

Using the shifting operator $S_{q}=e^{\frac{\lambda_{q}^{'}}{\omega_{q}}(b_{q}-b_{q}^{\dag})}$ above expression can be written as
\begin{equation}
H_{q}-\lambda_{q}^{'}(b_{q}+b_{q}^{\dag})=S_{q}H_{q}S_{q}^{\dag}-\frac{2\lambda_{q}'^{2}}{\omega_{q}},
\end{equation}
this implies
\begin{equation}
\dot{\tilde{U}}_{q}(t)=i\left(S_{q}H_{q}S_{q}^{\dag}-\frac{2\lambda_{q}'^{2}}{\omega_{q}}\right)\tilde{U}_{q}(t).
\end{equation}
Multiplying by $S_{q}^{\dag}$ on both sides
\begin{equation}
\frac{d(S_{q}^{\dag}\tilde{U}_{q}(t))}{dt}=i\left(H_{q}-\frac{2\lambda_{q}'^{2}}{\omega_{q}}\right)S_{q}^{\dag}\tilde{U}_{q}(t),
\end{equation}
solving for $S_{q}^{\dag}\tilde{U}_{q}(t)$ 
\begin{equation}
S_{q}^{\dag}\tilde{U}_{q}(t)=e^{iH_{q}t}S_{q}^{\dag}\tilde{U}_{q}(0)e^{-i\Delta_{P,q}t};\quad \Delta_{P,q}=\frac{2\lambda_{q}^{'2}}{\omega_{q}},
\end{equation}
using $U_{q}(t)=e^{-iH_{q}t}\tilde{U}_{q}(t)$ and $U_{q}(0)=1$, we obtain
\begin{equation}
U_{q}(t)=e^{-iH_{q}t}S_{q}e^{iH_{q}t}S_{q}^{\dag}e^{-i\Delta_{P,q}t}.
\end{equation}

Similarly, for $W_{q}(t)$
\begin{equation}
W_{q}(t)=S_{q}e^{-iH_{q}t}S_{q}^{\dag}e^{iH_{q}t}e^{i\Delta_{P,q}t}.
\end{equation}
Substituting for $U_{q}(t)$ and $W_{q}(t)$, also using $U_{q}(0)=W_{q}(0)=1$, we obtain an expression for $X_{P,q}(t)$
\begin{equation}
X_{P,q}(t)=S_{q}e^{iH_{q}t}S_{q}^{\dag}X_{P,q}(0)S_{q}e^{-iH_{q}t}S_{q}^{\dag}.
\end{equation}
Recalling Eq.\ (\ref{eqn:eom in ques}):
\begin{equation*}
\mathcal{C}(t)=2\underbrace{\langle X_{P}X_{P}(t)\rangle}_{\mathcal{C}_{1}(t)}+2\underbrace{\langle X_{P}(t)X_{P}\rangle}_{\mathcal{C}_{2}(t)}.
\end{equation*}
and substituting for $X_{P,q}(t)$, we have
\begin{align}
\mathcal{C}_{1}(t)=&\mathrm{Tr}_{P}\biggl[\sum_{q',q'''}X_{P,q'}(0)\biggl(\prod_{q''}S_{q''}e^{iH_{q''}t}S_{q''}^{\dag}\tilde{X}_{P,q'''}(0)\nonumber\\
&\times S_{q''}e^{-iH_{q''}t}S_{q''}^{\dag}\biggr)\biggl(\prod_{q}\rho_{P,q}(0)\biggr)\biggr].
\end{align}
Above expression can be solved for $q=q'=q''=q'''$, because other modes do not contribution in the final trace and do not conserve the particle number for the different modes. On further simplification and doing some algebraic manipulation, one can obtain a simplified expression as
\begin{widetext}
\begin{equation}
\mathcal{C}_{1}(t)=\sum_{q'=q'''}\prod_{q=q'=q''}\mathrm{Tr}_{P}\biggl[X_{P,q'}(0)\biggl(S_{q''}e^{iH_{q''}t}S_{q''}^{\dag}\tilde{X}_{P,q'''}(0)S_{q''}e^{-iH_{q''}t}S_{q''}^{\dag}\biggr)\rho_{P,q}(0)\biggr]
\end{equation}
and similarly for $\mathcal{C}_{2}(t)$
\begin{equation}
\mathcal{C}_{2}(t)=\sum_{q'=q'''}\prod_{q=q'=q''}\mathrm{Tr}_{P}\biggl[\biggl(S_{q''}e^{iH_{q''}t}S_{q''}^{\dag}\tilde{X}_{P,q'''}(0)S_{q''}e^{-iH_{q''}t}S_{q''}^{\dag}\biggr)X_{P,q'}(0)\rho_{P,q}(0)\biggr].
\end{equation}
\end{widetext}
Considering the factor in parenthesis which is common in both expressions, we have
\begin{equation}
X_{P}(t)=\sum_{q=q'}\prod_{q}\biggl(S_{q}\overbrace{e^{iH_{q}t}\underbrace{S_{q}^{\dag}\tilde{X}_{P,q'}(0)S_{q}}_{Z_{q}}e^{-iH_{q}t}}^{Y_{q}(t)}S_{q}^{\dag}\biggr),
\end{equation}
where we have defined:
\begin{equation*}
Y_{q}(t)=e^{iH_{q}t}Z_{q}e^{-iH_{q}t} \quad\mathrm{and} \qquad Z_{q}=S_{q}^{\dag}\tilde{X}_{P,q}(0)S_{q}.
\end{equation*}
Using Baker-Campbell-Hausdorff formula:
\begin{align}
Z_{q}&=S_{q}^{\dag}\tilde{X}_{P,q}(0)S_{q}\nonumber\\
&=\lambda_{q}'(b_{q}+b_{q}^{\dag})-\frac{2\lambda_{q}'^{2}}{\omega_{q}},
\end{align}
similarly for $Y_{q}(t)$,
\begin{align}
Y_{q}(t)&=e^{iH_{q}t}Z_{q}e^{-iH_{q}t}\nonumber\\
&=\lambda_{q}'(b_{q}e^{-i\omega_{q}t}+b_{q}^{\dag}e^{i\omega_{q}t})-\frac{2\lambda_{q}'^{2}}{\omega_{q}}.
\end{align}
Substituting for $Y_{q}(t)$ and $Z_{q}$ in the expression for $X_{P}(t)$, and then substituting for $X_{P}(t)$ in the expressions for $\mathcal{C}_{1}(t)$ and $\mathcal{C}_{2}(t)$, we perform the final trace to obtain the following expressions
\begin{align}
\mathcal{C}_{1}(t)&=\frac{\Delta}{4\Omega_{R}}\sum_{q}\lambda_{q}^{2}(n_{q}\,e^{i\omega_{q}t}+(n_{q}+1)e^{-i\omega_{q}t})\\
\mathcal{C}_{2}(t)&=\frac{\Delta}{4\Omega_{R}}\sum_{q}\lambda_{q}^{2}(n_{q}\,e^{-i\omega_{q}t}+(n_{q}+1)e^{i\omega_{q}t}).
\end{align}
After substituting the expressions for $\mathcal{C}_{1}(t)$ and $\mathcal{C}_{2}(t)$ in the expression for reduced self-energy matrix element and performing the final trace, we obtain
\begin{equation}
\Sigma_{+-,+-}^{P}(t)=\frac{-i\Delta^{2}\,e^{-i\Omega_{R}'t}}{2\Omega_{R}^{2}}\sum_{q}\lambda_{q}^{2}(2n_{q}+1)\,\cos(\omega_{q}t).
\end{equation}
Applying the continuum of modes and going back to Laplace domain,
\begin{align}
\Sigma_{+-,+-}^{P}(s)=&\frac{-i\alpha_{P}\Delta^{2}}{2\Omega_{R}^{2}}\int_{0}^{\infty}d\epsilon|\epsilon|^{3}e^{\frac{-|\epsilon|}{\epsilon_{c}}}(2n_{B}(\epsilon)+1)\nonumber\\
&\times\left(\frac{1}{s+i(\Omega_{R}'-\epsilon)}+\frac{1}{s+i(\Omega_{R}'+\epsilon)}\right),
\end{align}
above expression recovers the result obtained for self-energy matrix element within Born approximation, which is Eq.~(\ref{eqn:phonon:self:energy}) in the main text.

\section{\label{sec:initial-condition}Initial Condition}
In this section, we will discuss the stationary density matrix and find the initial condition for the operator $\Omega(t)$ given in Eq.~(\ref{eqn:corr:initial:condition}). The stationary density matrix $\bar{\rho}$ accounts for conditions that accumulate between the system and interactions in the time interval $t\in[t_{0} ,0]$. In this time interval, $[t_{0}, 0]$, the stationary density matrix is given by its value at $t=0$ and we can replace $\rho_{S}(t')\rightarrow\rho_{S}(t)$ in Eq.~(\ref{eqn:eom:rhos}), to obtain 
\begin{equation}
\dot{\rho}_{S}(t)=-iL_{S}\rho_{S}(t)-i\rho_{S}(t)\int_{t_{0}}^{t}dt'\Sigma_{S}(t-t').
\end{equation}
Changing integration variable: $\tau=t-t'$ to obtain
\begin{equation}
\dot{\rho}_{S}(t)=-iL_{S}\rho_{S}(t)-i\rho_{S}(t)\int_{0}^{t-t_{0}}d\tau\Sigma_{S}(\tau),
\end{equation}
one can extend the upper limit of above integration to infinity by setting $t_{0}\to-\infty$, and solving the differential equation to obtain
\begin{equation}
\rho_{S}(0)=e^{i[L_{S}+\Sigma_{S}(s=0)]t_0}\rho_{S}(t_0),
\end{equation}
and taking Laplace transform on both sides, we obtain
\begin{equation}
\frac{\rho_{S}(0)}{s}=\frac{1}{s+iL_{S}+i\Sigma_{S}(s=0)}\rho_{S}(t_{0}),
\end{equation}
here Laplace transform is defined as $f(s)=\int_{0}^{\infty}e^{-st}f(t)dt$. We choose an initial condition when exciton is in excited state $\ket{a}$ given by $\rho_{S}(t_{0})=\ket{a}\bra{a}$, and evolves in the presence of pump laser. After performing a secular approximation and using the definition of stationary limit
\begin{equation}
\bar{\rho}_{S}=\lim\limits_{s\rightarrow 0^{+}}s\left(\frac{\rho_{S}(0)}{s}\right),
\end{equation}
one can find all the elements of stationary density matrix operator as:
\begin{equation}
\bar{\rho}_{++}=\frac{-\Sigma_{++,--}(s=0)}{\Sigma_{z}(s=0)},
\end{equation}
where $\Sigma_{z}=\Sigma_{++,++}-\Sigma_{++,--}$. Similarly,
\begin{align}
\bar{\rho}_{--}&=\frac{\Sigma_{++,++}(s=0)}{\Sigma_{z}(s=0)}\\
\bar{\rho}_{+-}&=\bar{\rho}_{-+}=0.
\end{align}
Recalling, the initial condition for operator $\Omega(t)$ given by Eq.~(\ref{eqn:corr:initial:condition}):
\begin{equation}
\Omega(0)=\bar{\rho}\sigma_{ab},
\label{eqn:initial:condition:matrix}
\end{equation}
and matrix element of interest can be extracted after performing the trace over phonon and photon modes, and substituting for stationary density matrix element $\bar{\rho}_{++}$,
\begin{equation}
\Omega_{+-}(0)=\frac{-{\bf c}^{2}\Sigma_{++,--}(s=0)}{\Sigma_{z}(s=0)}.
\label{eqn:initial:condition:matrix:element}
\end{equation}
Dynamics of the operator $\Omega(t)$ is evaluated using Hamiltonian given by Eq.~(\ref{eqn:polaron:hamiltonian}), with an initial condition given by above expression which is Eq.~(\ref{eqn:initial:condition}) in the main text.
%\bibliography{tls}

\begin{thebibliography}{43}%
\makeatletter
\providecommand \@ifxundefined [1]{%
 \@ifx{#1\undefined}
}%
\providecommand \@ifnum [1]{%
 \ifnum #1\expandafter \@firstoftwo
 \else \expandafter \@secondoftwo
 \fi
}%
\providecommand \@ifx [1]{%
 \ifx #1\expandafter \@firstoftwo
 \else \expandafter \@secondoftwo
 \fi
}%
\providecommand \natexlab [1]{#1}%
\providecommand \enquote  [1]{``#1''}%
\providecommand \bibnamefont  [1]{#1}%
\providecommand \bibfnamefont [1]{#1}%
\providecommand \citenamefont [1]{#1}%
\providecommand \href@noop [0]{\@secondoftwo}%
\providecommand \href [0]{\begingroup \@sanitize@url \@href}%
\providecommand \@href[1]{\@@startlink{#1}\@@href}%
\providecommand \@@href[1]{\endgroup#1\@@endlink}%
\providecommand \@sanitize@url [0]{\catcode `\\12\catcode `\$12\catcode
  `\&12\catcode `\#12\catcode `\^12\catcode `\_12\catcode `\%12\relax}%
\providecommand \@@startlink[1]{}%
\providecommand \@@endlink[0]{}%
\providecommand \url  [0]{\begingroup\@sanitize@url \@url }%
\providecommand \@url [1]{\endgroup\@href {#1}{\urlprefix }}%
\providecommand \urlprefix  [0]{URL }%
\providecommand \Eprint [0]{\href }%
\providecommand \doibase [0]{http://dx.doi.org/}%
\providecommand \selectlanguage [0]{\@gobble}%
\providecommand \bibinfo  [0]{\@secondoftwo}%
\providecommand \bibfield  [0]{\@secondoftwo}%
\providecommand \translation [1]{[#1]}%
\providecommand \BibitemOpen [0]{}%
\providecommand \bibitemStop [0]{}%
\providecommand \bibitemNoStop [0]{.\EOS\space}%
\providecommand \EOS [0]{\spacefactor3000\relax}%
\providecommand \BibitemShut  [1]{\csname bibitem#1\endcsname}%
\let\auto@bib@innerbib\@empty
%</preamble>
\bibitem [{\citenamefont {von Neumann}(1955)}]{vonneumann55}%
  \BibitemOpen
  \bibfield  {author} {\bibinfo {author} {\bibfnamefont {J.}~\bibnamefont {von
  Neumann}},\ }\href@noop {} {\emph {\bibinfo {title} {Mathematical Foundations
  of Quantum Mechanics}}}\ (\bibinfo  {publisher} {Princeton University
  Press},\ \bibinfo {address} {Princeton, NJ, 195},\ \bibinfo {year}
  {1955})\BibitemShut {NoStop}%
\bibitem [{\citenamefont {Benioff}(1980)}]{benioff80}%
  \BibitemOpen
  \bibfield  {author} {\bibinfo {author} {\bibfnamefont {P.}~\bibnamefont
  {Benioff}},\ }\href@noop {} {\emph {\bibinfo {title} {J Stat Phys}}}\
  (\bibinfo {year} {1980})\BibitemShut {NoStop}%
\bibitem [{\citenamefont {Feynman}(1982)}]{feynman82}%
  \BibitemOpen
  \bibfield  {author} {\bibinfo {author} {\bibfnamefont {R.~P.}\ \bibnamefont
  {Feynman}},\ }\href@noop {} {\emph {\bibinfo {title} {Int J Theor Phys}}}\
  (\bibinfo {year} {1982})\BibitemShut {NoStop}%
\bibitem [{\citenamefont {Divincenzo}(1997)}]{loss97}%
  \BibitemOpen
  \bibfield  {author} {\bibinfo {author} {\bibfnamefont {D.~P.}\ \bibnamefont
  {Divincenzo}},\ }\enquote {\bibinfo {title} {Mesoscopic electron
  transport},}\ \ (\bibinfo  {publisher} {Springer Netherlands},\ \bibinfo
  {address} {Dordrecht},\ \bibinfo {year} {1997})\ Chap.\ \bibinfo {chapter}
  {Topics in Quantum Computers}, pp.\ \bibinfo {pages} {657--677}\BibitemShut
  {NoStop}%
\bibitem [{\citenamefont {Schumacher}(1995)}]{schumacher95}%
  \BibitemOpen
  \bibfield  {author} {\bibinfo {author} {\bibfnamefont {B.}~\bibnamefont
  {Schumacher}},\ }\href {\doibase 10.1103/PhysRevA.51.2738} {\bibfield
  {journal} {\bibinfo  {journal} {Phys. Rev. A}\ }\textbf {\bibinfo {volume}
  {51}},\ \bibinfo {pages} {2738} (\bibinfo {year} {1995})}\BibitemShut
  {NoStop}%
\bibitem [{\citenamefont {Loss}\ and\ \citenamefont
  {DiVincenzo}(1998)}]{loss98}%
  \BibitemOpen
  \bibfield  {author} {\bibinfo {author} {\bibfnamefont {D.}~\bibnamefont
  {Loss}}\ and\ \bibinfo {author} {\bibfnamefont {D.~P.}\ \bibnamefont
  {DiVincenzo}},\ }\href {\doibase 10.1103/PhysRevA.57.120} {\bibfield
  {journal} {\bibinfo  {journal} {Phys. Rev. A}\ }\textbf {\bibinfo {volume}
  {57}},\ \bibinfo {pages} {120} (\bibinfo {year} {1998})}\BibitemShut
  {NoStop}%
\bibitem [{\citenamefont {Zrenner}\ \emph {et~al.}(2002)\citenamefont
  {Zrenner}, \citenamefont {Beham}, \citenamefont {Stufler}, \citenamefont
  {Findeis}, \citenamefont {Bichler},\ and\ \citenamefont
  {Abstreiter}}]{zrenner02}%
  \BibitemOpen
  \bibfield  {author} {\bibinfo {author} {\bibfnamefont {A.}~\bibnamefont
  {Zrenner}}, \bibinfo {author} {\bibfnamefont {E.}~\bibnamefont {Beham}},
  \bibinfo {author} {\bibfnamefont {S.}~\bibnamefont {Stufler}}, \bibinfo
  {author} {\bibfnamefont {F.}~\bibnamefont {Findeis}}, \bibinfo {author}
  {\bibfnamefont {M.}~\bibnamefont {Bichler}}, \ and\ \bibinfo {author}
  {\bibfnamefont {G.}~\bibnamefont {Abstreiter}},\ }\href {\doibase
  10.1038/nature00912} {\bibfield  {journal} {\bibinfo  {journal} {Nature}\
  }\textbf {\bibinfo {volume} {418}} (\bibinfo {year} {2002}),\
  10.1038/nature00912}\BibitemShut {NoStop}%
\bibitem [{\citenamefont {Brunner}\ \emph {et~al.}(2009)\citenamefont
  {Brunner}, \citenamefont {Gerardot}, \citenamefont {Dalgarno}, \citenamefont
  {Wüst}, \citenamefont {Karrai}, \citenamefont {Stoltz}, \citenamefont
  {Petroff},\ and\ \citenamefont {Warburton}}]{brunner09}%
  \BibitemOpen
  \bibfield  {author} {\bibinfo {author} {\bibfnamefont {D.}~\bibnamefont
  {Brunner}}, \bibinfo {author} {\bibfnamefont {B.~D.}\ \bibnamefont
  {Gerardot}}, \bibinfo {author} {\bibfnamefont {P.~A.}\ \bibnamefont
  {Dalgarno}}, \bibinfo {author} {\bibfnamefont {G.}~\bibnamefont {Wüst}},
  \bibinfo {author} {\bibfnamefont {K.}~\bibnamefont {Karrai}}, \bibinfo
  {author} {\bibfnamefont {N.~G.}\ \bibnamefont {Stoltz}}, \bibinfo {author}
  {\bibfnamefont {P.~M.}\ \bibnamefont {Petroff}}, \ and\ \bibinfo {author}
  {\bibfnamefont {R.~J.}\ \bibnamefont {Warburton}},\ }\href {\doibase
  10.1126/science.1173684} {\bibfield  {journal} {\bibinfo  {journal}
  {Science}\ }\textbf {\bibinfo {volume} {325}},\ \bibinfo {pages} {70}
  (\bibinfo {year} {2009})}\BibitemShut {NoStop}%
\bibitem [{\citenamefont {Strau\ss{}}\ \emph {et~al.}(2016)\citenamefont
  {Strau\ss{}}, \citenamefont {Placke}, \citenamefont {Kreinberg},
  \citenamefont {Schneider}, \citenamefont {Kamp}, \citenamefont {H\"ofling},
  \citenamefont {Wolters},\ and\ \citenamefont {Reitzenstein}}]{strauss17}%
  \BibitemOpen
  \bibfield  {author} {\bibinfo {author} {\bibfnamefont {M.}~\bibnamefont
  {Strau\ss{}}}, \bibinfo {author} {\bibfnamefont {M.}~\bibnamefont {Placke}},
  \bibinfo {author} {\bibfnamefont {S.}~\bibnamefont {Kreinberg}}, \bibinfo
  {author} {\bibfnamefont {C.}~\bibnamefont {Schneider}}, \bibinfo {author}
  {\bibfnamefont {M.}~\bibnamefont {Kamp}}, \bibinfo {author} {\bibfnamefont
  {S.}~\bibnamefont {H\"ofling}}, \bibinfo {author} {\bibfnamefont
  {J.}~\bibnamefont {Wolters}}, \ and\ \bibinfo {author} {\bibfnamefont
  {S.}~\bibnamefont {Reitzenstein}},\ }\href {\doibase
  10.1103/PhysRevB.93.241306} {\bibfield  {journal} {\bibinfo  {journal} {Phys.
  Rev. B}\ }\textbf {\bibinfo {volume} {93}},\ \bibinfo {pages} {241306}
  (\bibinfo {year} {2016})}\BibitemShut {NoStop}%
\bibitem [{\citenamefont {Kuhlmann}\ \emph {et~al.}(2015)\citenamefont
  {Kuhlmann}, \citenamefont {Prechtel}, \citenamefont {Houel}, \citenamefont
  {Ludwig}, \citenamefont {Reuter}, \citenamefont {Wieck},\ and\ \citenamefont
  {Warburton}}]{warburton15}%
  \BibitemOpen
  \bibfield  {author} {\bibinfo {author} {\bibfnamefont {A.~V.}\ \bibnamefont
  {Kuhlmann}}, \bibinfo {author} {\bibfnamefont {J.~H.}\ \bibnamefont
  {Prechtel}}, \bibinfo {author} {\bibfnamefont {J.}~\bibnamefont {Houel}},
  \bibinfo {author} {\bibfnamefont {A.}~\bibnamefont {Ludwig}}, \bibinfo
  {author} {\bibfnamefont {D.}~\bibnamefont {Reuter}}, \bibinfo {author}
  {\bibfnamefont {A.~D.}\ \bibnamefont {Wieck}}, \ and\ \bibinfo {author}
  {\bibfnamefont {R.~J.}\ \bibnamefont {Warburton}},\ }\href {\doibase
  10.1038/ncomms9204} {\bibfield  {journal} {\bibinfo  {journal} {Nature}\
  }\textbf {\bibinfo {volume} {6}} (\bibinfo {year} {2015}),\
  10.1038/ncomms9204}\BibitemShut {NoStop}%
\bibitem [{\citenamefont {Ulhaq}\ \emph {et~al.}(2013)\citenamefont {Ulhaq},
  \citenamefont {Weiler}, \citenamefont {Roy}, \citenamefont {Ulrich},
  \citenamefont {Jetter}, \citenamefont {Hughes},\ and\ \citenamefont
  {Michler}}]{Ulhaq13}%
  \BibitemOpen
  \bibfield  {author} {\bibinfo {author} {\bibfnamefont {A.}~\bibnamefont
  {Ulhaq}}, \bibinfo {author} {\bibfnamefont {S.}~\bibnamefont {Weiler}},
  \bibinfo {author} {\bibfnamefont {C.}~\bibnamefont {Roy}}, \bibinfo {author}
  {\bibfnamefont {S.~M.}\ \bibnamefont {Ulrich}}, \bibinfo {author}
  {\bibfnamefont {M.}~\bibnamefont {Jetter}}, \bibinfo {author} {\bibfnamefont
  {S.}~\bibnamefont {Hughes}}, \ and\ \bibinfo {author} {\bibfnamefont
  {P.}~\bibnamefont {Michler}},\ }\href {\doibase 10.1364/OE.21.004382}
  {\bibfield  {journal} {\bibinfo  {journal} {Opt. Express}\ }\textbf {\bibinfo
  {volume} {21}},\ \bibinfo {pages} {4382} (\bibinfo {year}
  {2013})}\BibitemShut {NoStop}%
\bibitem [{\citenamefont {Cui}\ and\ \citenamefont {Raymer}(2006)}]{cui06}%
  \BibitemOpen
  \bibfield  {author} {\bibinfo {author} {\bibfnamefont {G.}~\bibnamefont
  {Cui}}\ and\ \bibinfo {author} {\bibfnamefont {M.~G.}\ \bibnamefont
  {Raymer}},\ }\href {\doibase 10.1103/PhysRevA.73.053807} {\bibfield
  {journal} {\bibinfo  {journal} {Phys. Rev. A}\ }\textbf {\bibinfo {volume}
  {73}},\ \bibinfo {pages} {053807} (\bibinfo {year} {2006})}\BibitemShut
  {NoStop}%
\bibitem [{\citenamefont {McCutcheon}\ and\ \citenamefont
  {Nazir}(2010)}]{dpsm10}%
  \BibitemOpen
  \bibfield  {author} {\bibinfo {author} {\bibfnamefont {D.~P.~S.}\
  \bibnamefont {McCutcheon}}\ and\ \bibinfo {author} {\bibfnamefont
  {A.}~\bibnamefont {Nazir}},\ }\href
  {http://stacks.iop.org/1367-2630/12/i=11/a=113042} {\bibfield  {journal}
  {\bibinfo  {journal} {New Journal of Physics}\ }\textbf {\bibinfo {volume}
  {12}},\ \bibinfo {pages} {113042} (\bibinfo {year} {2010})}\BibitemShut
  {NoStop}%
\bibitem [{\citenamefont {Ulrich}\ \emph {et~al.}(2011)\citenamefont {Ulrich},
  \citenamefont {Ates}, \citenamefont {Reitzenstein}, \citenamefont
  {L\"offler}, \citenamefont {Forchel},\ and\ \citenamefont
  {Michler}}]{ulrich11prl}%
  \BibitemOpen
  \bibfield  {author} {\bibinfo {author} {\bibfnamefont {S.~M.}\ \bibnamefont
  {Ulrich}}, \bibinfo {author} {\bibfnamefont {S.}~\bibnamefont {Ates}},
  \bibinfo {author} {\bibfnamefont {S.}~\bibnamefont {Reitzenstein}}, \bibinfo
  {author} {\bibfnamefont {A.}~\bibnamefont {L\"offler}}, \bibinfo {author}
  {\bibfnamefont {A.}~\bibnamefont {Forchel}}, \ and\ \bibinfo {author}
  {\bibfnamefont {P.}~\bibnamefont {Michler}},\ }\href {\doibase
  10.1103/PhysRevLett.106.247402} {\bibfield  {journal} {\bibinfo  {journal}
  {Phys. Rev. Lett.}\ }\textbf {\bibinfo {volume} {106}},\ \bibinfo {pages}
  {247402} (\bibinfo {year} {2011})}\BibitemShut {NoStop}%
\bibitem [{\citenamefont {Roy}\ and\ \citenamefont
  {Hughes}(2012)}]{hughes12prb}%
  \BibitemOpen
  \bibfield  {author} {\bibinfo {author} {\bibfnamefont {C.}~\bibnamefont
  {Roy}}\ and\ \bibinfo {author} {\bibfnamefont {S.}~\bibnamefont {Hughes}},\
  }\href {\doibase 10.1103/PhysRevB.85.115309} {\bibfield  {journal} {\bibinfo
  {journal} {Phys. Rev. B}\ }\textbf {\bibinfo {volume} {85}},\ \bibinfo
  {pages} {115309} (\bibinfo {year} {2012})}\BibitemShut {NoStop}%
\bibitem [{\citenamefont {McCutcheon}\ and\ \citenamefont
  {Nazir}(2013)}]{dpsm13prl}%
  \BibitemOpen
  \bibfield  {author} {\bibinfo {author} {\bibfnamefont {D.~P.~S.}\
  \bibnamefont {McCutcheon}}\ and\ \bibinfo {author} {\bibfnamefont
  {A.}~\bibnamefont {Nazir}},\ }\href {\doibase 10.1103/PhysRevLett.110.217401}
  {\bibfield  {journal} {\bibinfo  {journal} {Phys. Rev. Lett.}\ }\textbf
  {\bibinfo {volume} {110}},\ \bibinfo {pages} {217401} (\bibinfo {year}
  {2013})}\BibitemShut {NoStop}%
\bibitem [{\citenamefont {Weiler}\ \emph {et~al.}(2012)\citenamefont {Weiler},
  \citenamefont {Ulhaq}, \citenamefont {Ulrich}, \citenamefont {Richter},
  \citenamefont {Jetter}, \citenamefont {Michler}, \citenamefont {Roy},\ and\
  \citenamefont {Hughes}}]{weiler12}%
  \BibitemOpen
  \bibfield  {author} {\bibinfo {author} {\bibfnamefont {S.}~\bibnamefont
  {Weiler}}, \bibinfo {author} {\bibfnamefont {A.}~\bibnamefont {Ulhaq}},
  \bibinfo {author} {\bibfnamefont {S.~M.}\ \bibnamefont {Ulrich}}, \bibinfo
  {author} {\bibfnamefont {D.}~\bibnamefont {Richter}}, \bibinfo {author}
  {\bibfnamefont {M.}~\bibnamefont {Jetter}}, \bibinfo {author} {\bibfnamefont
  {P.}~\bibnamefont {Michler}}, \bibinfo {author} {\bibfnamefont
  {C.}~\bibnamefont {Roy}}, \ and\ \bibinfo {author} {\bibfnamefont
  {S.}~\bibnamefont {Hughes}},\ }\href {\doibase 10.1103/PhysRevB.86.241304}
  {\bibfield  {journal} {\bibinfo  {journal} {Phys. Rev. B}\ }\textbf {\bibinfo
  {volume} {86}},\ \bibinfo {pages} {241304} (\bibinfo {year}
  {2012})}\BibitemShut {NoStop}%
\bibitem [{\citenamefont {Krummheuer}\ \emph {et~al.}(2002)\citenamefont
  {Krummheuer}, \citenamefont {Axt},\ and\ \citenamefont
  {Kuhn}}]{krummheuer02}%
  \BibitemOpen
  \bibfield  {author} {\bibinfo {author} {\bibfnamefont {B.}~\bibnamefont
  {Krummheuer}}, \bibinfo {author} {\bibfnamefont {V.~M.}\ \bibnamefont {Axt}},
  \ and\ \bibinfo {author} {\bibfnamefont {T.}~\bibnamefont {Kuhn}},\ }\href
  {\doibase 10.1103/PhysRevB.65.195313} {\bibfield  {journal} {\bibinfo
  {journal} {Phys. Rev. B}\ }\textbf {\bibinfo {volume} {65}},\ \bibinfo
  {pages} {195313} (\bibinfo {year} {2002})}\BibitemShut {NoStop}%
\bibitem [{\citenamefont {Vagov}\ \emph {et~al.}(2014)\citenamefont {Vagov},
  \citenamefont {Gl\"assl}, \citenamefont {Croitoru}, \citenamefont {Axt},\
  and\ \citenamefont {Kuhn}}]{vagov14}%
  \BibitemOpen
  \bibfield  {author} {\bibinfo {author} {\bibfnamefont {A.}~\bibnamefont
  {Vagov}}, \bibinfo {author} {\bibfnamefont {M.}~\bibnamefont {Gl\"assl}},
  \bibinfo {author} {\bibfnamefont {M.~D.}\ \bibnamefont {Croitoru}}, \bibinfo
  {author} {\bibfnamefont {V.~M.}\ \bibnamefont {Axt}}, \ and\ \bibinfo
  {author} {\bibfnamefont {T.}~\bibnamefont {Kuhn}},\ }\href {\doibase
  10.1103/PhysRevB.90.075309} {\bibfield  {journal} {\bibinfo  {journal} {Phys.
  Rev. B}\ }\textbf {\bibinfo {volume} {90}},\ \bibinfo {pages} {075309}
  (\bibinfo {year} {2014})}\BibitemShut {NoStop}%
\bibitem [{\citenamefont {Hughes}\ and\ \citenamefont
  {Agarwal}(2017)}]{agarwal17}%
  \BibitemOpen
  \bibfield  {author} {\bibinfo {author} {\bibfnamefont {S.}~\bibnamefont
  {Hughes}}\ and\ \bibinfo {author} {\bibfnamefont {G.~S.}\ \bibnamefont
  {Agarwal}},\ }\href {\doibase 10.1103/PhysRevLett.118.063601} {\bibfield
  {journal} {\bibinfo  {journal} {Phys. Rev. Lett.}\ }\textbf {\bibinfo
  {volume} {118}},\ \bibinfo {pages} {063601} (\bibinfo {year}
  {2017})}\BibitemShut {NoStop}%
\bibitem [{\citenamefont {Kryuchkyan}\ \emph {et~al.}(2017)\citenamefont
  {Kryuchkyan}, \citenamefont {Shahnazaryan}, \citenamefont {Kibis},\ and\
  \citenamefont {Shelykh}}]{shelykh17}%
  \BibitemOpen
  \bibfield  {author} {\bibinfo {author} {\bibfnamefont {G.~Y.}\ \bibnamefont
  {Kryuchkyan}}, \bibinfo {author} {\bibfnamefont {V.}~\bibnamefont
  {Shahnazaryan}}, \bibinfo {author} {\bibfnamefont {O.~V.}\ \bibnamefont
  {Kibis}}, \ and\ \bibinfo {author} {\bibfnamefont {I.~A.}\ \bibnamefont
  {Shelykh}},\ }\href {\doibase 10.1103/PhysRevA.95.013834} {\bibfield
  {journal} {\bibinfo  {journal} {Phys. Rev. A}\ }\textbf {\bibinfo {volume}
  {95}},\ \bibinfo {pages} {013834} (\bibinfo {year} {2017})}\BibitemShut
  {NoStop}%
\bibitem [{\citenamefont {Coish}\ and\ \citenamefont {Loss}(2004)}]{bill04}%
  \BibitemOpen
  \bibfield  {author} {\bibinfo {author} {\bibfnamefont {W.~A.}\ \bibnamefont
  {Coish}}\ and\ \bibinfo {author} {\bibfnamefont {D.}~\bibnamefont {Loss}},\
  }\href {\doibase 10.1103/PhysRevB.70.195340} {\bibfield  {journal} {\bibinfo
  {journal} {Phys. Rev. B}\ }\textbf {\bibinfo {volume} {70}},\ \bibinfo
  {pages} {195340} (\bibinfo {year} {2004})}\BibitemShut {NoStop}%
\bibitem [{\citenamefont {Kaer}\ \emph {et~al.}(2010)\citenamefont {Kaer},
  \citenamefont {Nielsen}, \citenamefont {Lodahl}, \citenamefont {Jauho},\ and\
  \citenamefont {M\o{}rk}}]{kaer10}%
  \BibitemOpen
  \bibfield  {author} {\bibinfo {author} {\bibfnamefont {P.}~\bibnamefont
  {Kaer}}, \bibinfo {author} {\bibfnamefont {T.~R.}\ \bibnamefont {Nielsen}},
  \bibinfo {author} {\bibfnamefont {P.}~\bibnamefont {Lodahl}}, \bibinfo
  {author} {\bibfnamefont {A.-P.}\ \bibnamefont {Jauho}}, \ and\ \bibinfo
  {author} {\bibfnamefont {J.}~\bibnamefont {M\o{}rk}},\ }\href {\doibase
  10.1103/PhysRevLett.104.157401} {\bibfield  {journal} {\bibinfo  {journal}
  {Phys. Rev. Lett.}\ }\textbf {\bibinfo {volume} {104}},\ \bibinfo {pages}
  {157401} (\bibinfo {year} {2010})}\BibitemShut {NoStop}%
\bibitem [{\citenamefont {Swain}(1981)}]{swan81}%
  \BibitemOpen
  \bibfield  {author} {\bibinfo {author} {\bibfnamefont {S.}~\bibnamefont
  {Swain}},\ }\href@noop {} {\bibfield  {journal} {\bibinfo  {journal} {Journal
  of Physics A: Mathematical and General}\ }\textbf {\bibinfo {volume} {14}},\
  \bibinfo {pages} {2577} (\bibinfo {year} {1981})}\BibitemShut {NoStop}%
\bibitem [{\citenamefont {Thanopulos}\ \emph {et~al.}(2017)\citenamefont
  {Thanopulos}, \citenamefont {Yannopapas},\ and\ \citenamefont
  {Paspalakis}}]{thanopulos17}%
  \BibitemOpen
  \bibfield  {author} {\bibinfo {author} {\bibfnamefont {I.}~\bibnamefont
  {Thanopulos}}, \bibinfo {author} {\bibfnamefont {V.}~\bibnamefont
  {Yannopapas}}, \ and\ \bibinfo {author} {\bibfnamefont {E.}~\bibnamefont
  {Paspalakis}},\ }\href {\doibase 10.1103/PhysRevB.95.075412} {\bibfield
  {journal} {\bibinfo  {journal} {Phys. Rev. B}\ }\textbf {\bibinfo {volume}
  {95}},\ \bibinfo {pages} {075412} (\bibinfo {year} {2017})}\BibitemShut
  {NoStop}%
\bibitem [{\citenamefont {Toyli}\ \emph {et~al.}(2016)\citenamefont {Toyli},
  \citenamefont {Eddins}, \citenamefont {Boutin}, \citenamefont {Puri},
  \citenamefont {Hover}, \citenamefont {Bolkhovsky}, \citenamefont {Oliver},
  \citenamefont {Blais},\ and\ \citenamefont {Siddiqi}}]{toyli16}%
  \BibitemOpen
  \bibfield  {author} {\bibinfo {author} {\bibfnamefont {D.~M.}\ \bibnamefont
  {Toyli}}, \bibinfo {author} {\bibfnamefont {A.~W.}\ \bibnamefont {Eddins}},
  \bibinfo {author} {\bibfnamefont {S.}~\bibnamefont {Boutin}}, \bibinfo
  {author} {\bibfnamefont {S.}~\bibnamefont {Puri}}, \bibinfo {author}
  {\bibfnamefont {D.}~\bibnamefont {Hover}}, \bibinfo {author} {\bibfnamefont
  {V.}~\bibnamefont {Bolkhovsky}}, \bibinfo {author} {\bibfnamefont {W.~D.}\
  \bibnamefont {Oliver}}, \bibinfo {author} {\bibfnamefont {A.}~\bibnamefont
  {Blais}}, \ and\ \bibinfo {author} {\bibfnamefont {I.}~\bibnamefont
  {Siddiqi}},\ }\href {\doibase 10.1103/PhysRevX.6.031004} {\bibfield
  {journal} {\bibinfo  {journal} {Phys. Rev. X}\ }\textbf {\bibinfo {volume}
  {6}},\ \bibinfo {pages} {031004} (\bibinfo {year} {2016})}\BibitemShut
  {NoStop}%
\bibitem [{\citenamefont {McCutcheon}(2016)}]{dpsm16}%
  \BibitemOpen
  \bibfield  {author} {\bibinfo {author} {\bibfnamefont {D.~P.~S.}\
  \bibnamefont {McCutcheon}},\ }\href {\doibase 10.1103/PhysRevA.93.022119}
  {\bibfield  {journal} {\bibinfo  {journal} {Phys. Rev. A}\ }\textbf {\bibinfo
  {volume} {93}},\ \bibinfo {pages} {022119} (\bibinfo {year}
  {2016})}\BibitemShut {NoStop}%
\bibitem [{\citenamefont {Fick}\ and\ \citenamefont {Sauermann}(1990)}]{fick}%
  \BibitemOpen
  \bibfield  {author} {\bibinfo {author} {\bibfnamefont {E.}~\bibnamefont
  {Fick}}\ and\ \bibinfo {author} {\bibfnamefont {G.}~\bibnamefont
  {Sauermann}},\ }\href@noop {} {\emph {\bibinfo {title} {The Quantum
  Statistics of Dynamic Processes}}}\ (\bibinfo  {publisher}
  {Springer-Verlag},\ \bibinfo {address} {Berlin},\ \bibinfo {year}
  {1990})\BibitemShut {NoStop}%
\bibitem [{\citenamefont {Muller}\ \emph {et~al.}(2007)\citenamefont {Muller},
  \citenamefont {Flagg}, \citenamefont {Bianucci}, \citenamefont {Wang},
  \citenamefont {Deppe}, \citenamefont {Ma}, \citenamefont {Zhang},
  \citenamefont {Salamo}, \citenamefont {Xiao},\ and\ \citenamefont
  {Shih}}]{muller07}%
  \BibitemOpen
  \bibfield  {author} {\bibinfo {author} {\bibfnamefont {A.}~\bibnamefont
  {Muller}}, \bibinfo {author} {\bibfnamefont {E.~B.}\ \bibnamefont {Flagg}},
  \bibinfo {author} {\bibfnamefont {P.}~\bibnamefont {Bianucci}}, \bibinfo
  {author} {\bibfnamefont {X.~Y.}\ \bibnamefont {Wang}}, \bibinfo {author}
  {\bibfnamefont {D.~G.}\ \bibnamefont {Deppe}}, \bibinfo {author}
  {\bibfnamefont {W.}~\bibnamefont {Ma}}, \bibinfo {author} {\bibfnamefont
  {J.}~\bibnamefont {Zhang}}, \bibinfo {author} {\bibfnamefont {G.~J.}\
  \bibnamefont {Salamo}}, \bibinfo {author} {\bibfnamefont {M.}~\bibnamefont
  {Xiao}}, \ and\ \bibinfo {author} {\bibfnamefont {C.~K.}\ \bibnamefont
  {Shih}},\ }\href {\doibase 10.1103/PhysRevLett.99.187402} {\bibfield
  {journal} {\bibinfo  {journal} {Phys. Rev. Lett.}\ }\textbf {\bibinfo
  {volume} {99}},\ \bibinfo {pages} {187402} (\bibinfo {year}
  {2007})}\BibitemShut {NoStop}%
\bibitem [{\citenamefont {Mahan}(1990)}]{mahan}%
  \BibitemOpen
  \bibfield  {author} {\bibinfo {author} {\bibfnamefont {G.}~\bibnamefont
  {Mahan}},\ }\href@noop {} {\emph {\bibinfo {title} {Many-particle Physics}}}\
  (\bibinfo  {publisher} {Plenum},\ \bibinfo {address} {New York},\ \bibinfo
  {year} {1990})\BibitemShut {NoStop}%
\bibitem [{\citenamefont {Mollow}(1969)}]{Mollow69}%
  \BibitemOpen
  \bibfield  {author} {\bibinfo {author} {\bibfnamefont {B.~R.}\ \bibnamefont
  {Mollow}},\ }\href {\doibase 10.1103/PhysRev.188.1969} {\bibfield  {journal}
  {\bibinfo  {journal} {Phys. Rev.}\ }\textbf {\bibinfo {volume} {188}},\
  \bibinfo {pages} {1969} (\bibinfo {year} {1969})}\BibitemShut {NoStop}%
\bibitem [{\citenamefont {Cohen-Tannoudji}\ and\ \citenamefont
  {Grynberg}(2004)}]{cohen}%
  \BibitemOpen
  \bibfield  {author} {\bibinfo {author} {\bibfnamefont {J.~D.-R.}\
  \bibnamefont {Cohen-Tannoudji}}\ and\ \bibinfo {author} {\bibfnamefont
  {G.}~\bibnamefont {Grynberg}},\ }\href@noop {} {\emph {\bibinfo {title}
  {Atom-Photon Interactions}}}\ (\bibinfo  {publisher} {Wiley-VCH},\ \bibinfo
  {address} {Berlin},\ \bibinfo {year} {2004})\BibitemShut {NoStop}%
\bibitem [{\citenamefont {Hopfmann}\ \emph {et~al.}(2017)\citenamefont
  {Hopfmann}, \citenamefont {Carmele}, \citenamefont {Musia\l{}}, \citenamefont
  {Schneider}, \citenamefont {Kamp}, \citenamefont {H\"ofling}, \citenamefont
  {Knorr},\ and\ \citenamefont {Reitzenstein}}]{hopfmann17}%
  \BibitemOpen
  \bibfield  {author} {\bibinfo {author} {\bibfnamefont {C.}~\bibnamefont
  {Hopfmann}}, \bibinfo {author} {\bibfnamefont {A.}~\bibnamefont {Carmele}},
  \bibinfo {author} {\bibfnamefont {A.}~\bibnamefont {Musia\l{}}}, \bibinfo
  {author} {\bibfnamefont {C.}~\bibnamefont {Schneider}}, \bibinfo {author}
  {\bibfnamefont {M.}~\bibnamefont {Kamp}}, \bibinfo {author} {\bibfnamefont
  {S.}~\bibnamefont {H\"ofling}}, \bibinfo {author} {\bibfnamefont
  {A.}~\bibnamefont {Knorr}}, \ and\ \bibinfo {author} {\bibfnamefont
  {S.}~\bibnamefont {Reitzenstein}},\ }\href {\doibase
  10.1103/PhysRevB.95.035302} {\bibfield  {journal} {\bibinfo  {journal} {Phys.
  Rev. B}\ }\textbf {\bibinfo {volume} {95}},\ \bibinfo {pages} {035302}
  (\bibinfo {year} {2017})}\BibitemShut {NoStop}%
\bibitem [{\citenamefont {Laucht}\ \emph {et~al.}(2016)\citenamefont {Laucht},
  \citenamefont {Simmons}, \citenamefont {Kalra}, \citenamefont {Tosi},
  \citenamefont {Dehollain}, \citenamefont {Muhonen}, \citenamefont {Freer},
  \citenamefont {Hudson}, \citenamefont {Itoh}, \citenamefont {Jamieson},
  \citenamefont {McCallum}, \citenamefont {Dzurak},\ and\ \citenamefont
  {Morello}}]{laucht17}%
  \BibitemOpen
  \bibfield  {author} {\bibinfo {author} {\bibfnamefont {A.}~\bibnamefont
  {Laucht}}, \bibinfo {author} {\bibfnamefont {S.}~\bibnamefont {Simmons}},
  \bibinfo {author} {\bibfnamefont {R.}~\bibnamefont {Kalra}}, \bibinfo
  {author} {\bibfnamefont {G.}~\bibnamefont {Tosi}}, \bibinfo {author}
  {\bibfnamefont {J.~P.}\ \bibnamefont {Dehollain}}, \bibinfo {author}
  {\bibfnamefont {J.~T.}\ \bibnamefont {Muhonen}}, \bibinfo {author}
  {\bibfnamefont {S.}~\bibnamefont {Freer}}, \bibinfo {author} {\bibfnamefont
  {F.~E.}\ \bibnamefont {Hudson}}, \bibinfo {author} {\bibfnamefont {K.~M.}\
  \bibnamefont {Itoh}}, \bibinfo {author} {\bibfnamefont {D.~N.}\ \bibnamefont
  {Jamieson}}, \bibinfo {author} {\bibfnamefont {J.~C.}\ \bibnamefont
  {McCallum}}, \bibinfo {author} {\bibfnamefont {A.~S.}\ \bibnamefont
  {Dzurak}}, \ and\ \bibinfo {author} {\bibfnamefont {A.}~\bibnamefont
  {Morello}},\ }\href {\doibase 10.1103/PhysRevB.94.161302} {\bibfield
  {journal} {\bibinfo  {journal} {Phys. Rev. B}\ }\textbf {\bibinfo {volume}
  {94}},\ \bibinfo {pages} {161302} (\bibinfo {year} {2016})}\BibitemShut
  {NoStop}%
\bibitem [{\citenamefont {Scully}\ and\ \citenamefont
  {Zubairy}(1997)}]{scully}%
  \BibitemOpen
  \bibfield  {author} {\bibinfo {author} {\bibfnamefont {M.~O.}\ \bibnamefont
  {Scully}}\ and\ \bibinfo {author} {\bibfnamefont {M.~S.}\ \bibnamefont
  {Zubairy}},\ }\href@noop {} {\emph {\bibinfo {title} {Quantum Optics}}}\
  (\bibinfo  {publisher} {Cambridge University Press},\ \bibinfo {address}
  {Cambridge},\ \bibinfo {year} {1997})\BibitemShut {NoStop}%
\bibitem [{\citenamefont {de~Vega}\ and\ \citenamefont
  {Alonso}(2008)}]{vega08}%
  \BibitemOpen
  \bibfield  {author} {\bibinfo {author} {\bibfnamefont {I.}~\bibnamefont
  {de~Vega}}\ and\ \bibinfo {author} {\bibfnamefont {D.}~\bibnamefont
  {Alonso}},\ }\href {\doibase 10.1103/PhysRevA.77.043836} {\bibfield
  {journal} {\bibinfo  {journal} {Phys. Rev. A}\ }\textbf {\bibinfo {volume}
  {77}},\ \bibinfo {pages} {043836} (\bibinfo {year} {2008})}\BibitemShut
  {NoStop}%
\bibitem [{\citenamefont {Lax}(2000)}]{lax00}%
  \BibitemOpen
  \bibfield  {author} {\bibinfo {author} {\bibfnamefont {M.}~\bibnamefont
  {Lax}},\ }\href {\doibase http://dx.doi.org/10.1016/S0030-4018(00)00622-2}
  {\bibfield  {journal} {\bibinfo  {journal} {Optics Communications}\ }\textbf
  {\bibinfo {volume} {179}},\ \bibinfo {pages} {463 } (\bibinfo {year}
  {2000})}\BibitemShut {NoStop}%
\bibitem [{\citenamefont {Jin}\ \emph {et~al.}(2016)\citenamefont {Jin},
  \citenamefont {Karlewski},\ and\ \citenamefont {Marthaler}}]{jin16}%
  \BibitemOpen
  \bibfield  {author} {\bibinfo {author} {\bibfnamefont {J.}~\bibnamefont
  {Jin}}, \bibinfo {author} {\bibfnamefont {C.}~\bibnamefont {Karlewski}}, \
  and\ \bibinfo {author} {\bibfnamefont {M.}~\bibnamefont {Marthaler}},\ }\href
  {http://stacks.iop.org/1367-2630/18/i=8/a=083038} {\bibfield  {journal}
  {\bibinfo  {journal} {New Journal of Physics}\ }\textbf {\bibinfo {volume}
  {18}},\ \bibinfo {pages} {083038} (\bibinfo {year} {2016})}\BibitemShut
  {NoStop}%
\bibitem [{\citenamefont {M.E.}\ and\ \citenamefont {Rumyantsev}(1996)}]{qd1}%
  \BibitemOpen
  \bibfield  {author} {\bibinfo {author} {\bibfnamefont {L.}~\bibnamefont
  {M.E.}}\ and\ \bibinfo {author} {\bibfnamefont {S.}~\bibnamefont
  {Rumyantsev}},\ }\href@noop {} {\emph {\bibinfo {title} {Handbook Series on
  Semiconductor Parameters}}}\ (\bibinfo  {publisher} {World Scientific},\
  \bibinfo {address} {London},\ \bibinfo {year} {1996})\BibitemShut {NoStop}%
\bibitem [{\citenamefont {A.}\ and\ \citenamefont {Kundrotas}(1994)}]{qd2}%
  \BibitemOpen
  \bibfield  {author} {\bibinfo {author} {\bibfnamefont {D.}~\bibnamefont
  {A.}}\ and\ \bibinfo {author} {\bibfnamefont {J.}~\bibnamefont {Kundrotas}},\
  }\href@noop {} {\emph {\bibinfo {title} {Handbook on Physical Properties of
  Ge, Si, GaAs and InP}}}\ (\bibinfo  {publisher} {Science and Encyclopedia
  Publishers},\ \bibinfo {address} {Vilnius},\ \bibinfo {year}
  {1994})\BibitemShut {NoStop}%
\bibitem [{\citenamefont {Schrieffer}\ and\ \citenamefont {Wolff}(1966)}]{swt}%
  \BibitemOpen
  \bibfield  {author} {\bibinfo {author} {\bibfnamefont {J.~R.}\ \bibnamefont
  {Schrieffer}}\ and\ \bibinfo {author} {\bibfnamefont {P.~A.}\ \bibnamefont
  {Wolff}},\ }\href {\doibase 10.1103/PhysRev.149.491} {\bibfield  {journal}
  {\bibinfo  {journal} {Phys. Rev.}\ }\textbf {\bibinfo {volume} {149}},\
  \bibinfo {pages} {491} (\bibinfo {year} {1966})}\BibitemShut {NoStop}%
\bibitem [{\citenamefont {DiVincenzo}\ and\ \citenamefont
  {Loss}(2005)}]{loss05}%
  \BibitemOpen
  \bibfield  {author} {\bibinfo {author} {\bibfnamefont {D.~P.}\ \bibnamefont
  {DiVincenzo}}\ and\ \bibinfo {author} {\bibfnamefont {D.}~\bibnamefont
  {Loss}},\ }\href {\doibase 10.1103/PhysRevB.71.035318} {\bibfield  {journal}
  {\bibinfo  {journal} {Phys. Rev. B}\ }\textbf {\bibinfo {volume} {71}},\
  \bibinfo {pages} {035318} (\bibinfo {year} {2005})}\BibitemShut {NoStop}%
\bibitem [{\citenamefont {Ma}\ \emph {et~al.}(2012)\citenamefont {Ma},
  \citenamefont {Sun}, \citenamefont {Wang},\ and\ \citenamefont
  {Nori}}]{nori12}%
  \BibitemOpen
  \bibfield  {author} {\bibinfo {author} {\bibfnamefont {J.}~\bibnamefont
  {Ma}}, \bibinfo {author} {\bibfnamefont {Z.}~\bibnamefont {Sun}}, \bibinfo
  {author} {\bibfnamefont {X.}~\bibnamefont {Wang}}, \ and\ \bibinfo {author}
  {\bibfnamefont {F.}~\bibnamefont {Nori}},\ }\href {\doibase
  10.1103/PhysRevA.85.062323} {\bibfield  {journal} {\bibinfo  {journal} {Phys.
  Rev. A}\ }\textbf {\bibinfo {volume} {85}},\ \bibinfo {pages} {062323}
  (\bibinfo {year} {2012})}\BibitemShut {NoStop}%
\end{thebibliography}
\end{document}